\newcommand \Mpc {h^{-1}{\rm Mpc}}
\newcommand \kpc {h^{-1}{\rm kpc}}

\newcommand \arcm{\hbox{$^{\prime}$}}

\newcommand \kms {{\rm km~s}^{-1}}

\newcommand \beqn {\begin{equation}}
\newcommand \eeqn {\end{equation}}
\newcommand \ncirs {72 }

\documentclass[12pt,preprint]{aastex}
\usepackage{emulateapj5}

\begin{document}

\title{CIRS: Cluster Infall Regions in the Sloan Digital Sky Survey
I. Infall Patterns and Mass Profiles}
%\shorttitle{Cluster Infall Regions in SDSS}
%\shortauthors{Rines and Diaferio}

\author{Kenneth Rines\altaffilmark{1}
and Antonaldo Diaferio\altaffilmark{2}} 
\email{krines@astro.yale.edu}

\altaffiltext{1}{Yale Center for Astronomy and Astrophysics, Yale University, PO Box 208121, New Haven, CT 06520-8121 ; krines@astro.yale.edu}
\altaffiltext{2}{Universit\`a degli Studi di Torino,
Dipartimento di Fisica Generale ``Amedeo Avogadro'', Torino, Italy; diaferio@ph.unito.it}

\begin{abstract}

We use the Fourth Data Release of the Sloan Digital Sky Survey to test
the ubiquity of infall patterns around galaxy clusters and measure
cluster mass profiles to large radii.  The Cluster And Infall Region
Nearby Survey (CAIRNS) found infall patterns in nine clusters, but the
cluster sample was incomplete.  Here, we match X-ray cluster catalogs
with SDSS, search for infall patterns, and compute mass profiles for a
complete sample of X-ray selected clusters.  Very clean infall
patterns are apparent in most of the clusters, with the fraction
decreasing with increasing redshift due to shallower sampling.  All 72
clusters in a well-defined sample limited by redshift (ensuring good
sampling) and X-ray flux (excluding superpositions) show infall
patterns sufficient to apply the caustic technique.  This sample is by
far the largest sample of cluster mass profiles extending to large
radii to date.
%This high success rate is better than predicted by simulations.  
Similar to CAIRNS, cluster infall patterns are better defined in
observations than in simulations.  Further work is needed to determine
the source of this difference.  We use the infall patterns to compute
mass profiles for 72 clusters and compare them to model profiles.
Cluster scaling relations using caustic masses agree well with those
using X-ray or virial mass estimates, confirming the reliability of
the caustic technique.  We confirm the conclusion of CAIRNS that
cluster infall regions are well fit by NFW and Hernquist profiles and
poorly fit by singular isothermal spheres.  This much larger sample
enables new comparisons of cluster properties with those in
simulations.  The shapes (specifically, NFW concentrations) of the
mass profiles agree well with the predictions of simulations.  The
mass in the infall region is typically comparable to or larger than
that in the virial region.  Specifically, the mass inside the
turnaround radius is on average 2.19$\pm$0.18 times that within the
virial radius.  This ratio agrees well with recent predictions from
simulations of the final masses of dark matter haloes.

\end{abstract}

\keywords{galaxies: clusters: individual  --- galaxies: 
kinematics and dynamics --- cosmology: observations }

\section{Introduction}

Clusters of galaxies are the most massive gravitationally relaxed
systems in the universe.  They offer a unique probe of the properties
of galaxies and the distribution of matter on intermediate scales.
The dynamically relaxed centers of clusters are surrounded by infall
regions in which galaxies are bound to the cluster but are not in
equilibrium.  The Cluster and Infall Region Nearby Survey (CAIRNS)
pioneered the detailed study of cluster infall regions in
observations.  CAIRNS studied nine nearby galaxy clusters and their
infall regions with extensive spectroscopy \citep{cairnsi,cairnsha}
and near-infrared photometry from the Two-Micron All-Sky Survey
\citep[][]{cairnsii}.  The nine CAIRNS clusters display a
characteristic trumpet-shaped pattern in radius-redshift phase space
diagrams.  These patterns were first predicted for simple spherical
infall onto clusters \citep{kais87,rg89}, but later work showed that
these patterns reflect the dynamics of the infall region
\citep[][hereafter DG]{dg97} and \citep[][hereafter D99]{diaferio1999}.

Using numerical simulations, DG and D99 showed that the amplitude of
the caustics is a measure of the escape velocity from the cluster;
identification of the caustics therefore allows a determination of the
mass profile of the cluster on scales $\lesssim 10\Mpc$.  In
particular, nonparametric measurements of caustics yield cluster mass
profiles accurate to $\sim$50\% on scales of up to 10 $h^{-1}$ Mpc
when applied to clusters extracted from cosmological simulations (thus
accounting for many potential systematic uncertainties, primarily
departures from spherical symmetry).  This method assumes only that
galaxies trace the velocity field.  Many recent simulations suggest
that little ($\sim$10\%) or no velocity bias exists on linear and
mildly non-linear scales
\citep{kauffmann1999a,kauffmann1999b,diemand04,faltenbacher05}.
Velocity bias might be more significant in the centers of clusters,
where the galaxies have experienced dynamical friction.  Dynamical
friction should produce a smaller velocity dispersion for cluster
galaxies than dark matter
\citep{1999ApJ...516..530K,2003ApJ...590..654Y,2005MNRAS.361.1203V},
although they may undergo more frequent mergers (due to their smaller
velocities) or tidal disruption, resulting in a larger velocity
dispersion of the surviving cluster galaxies relative to the dark
matter \citep{1999ApJ...523...32C,diemand04,faltenbacher05}.  Note
also that the velocity bias may depend on the mass of the cluster
\citep{berlind03} or the luminosities of the tracer galaxies
\citep{2006MNRAS.366.1455S}.  An important caveat is that the subhalo
distribution of cluster galaxies in simulations does not match the
observed distributions of cluster galaxies: real cluster galaxies
trace the dark matter profile, while simulated cluster galaxies are
significantly antibiased in cluster centers, likely due to the
resilience of the stellar component of a subhalo against tidal
disruption relative to the total subhalo
\citep[e.g.,][]{diemand04,faltenbacher06}.  The details of velocity
bias depend also on the selection used to identify subhaloes, e.g.,
whether subhalo mass is measured prior to or after a subhalo enters
the cluster \citep{faltenbacher06}.  Because the mechanisms that may
produce velocity bias depend strongly on local density, the velocity
bias outside cluster cores (where the caustic technique is applied)
should be smaller.

CAIRNS showed that infall patterns are well defined in observations of
nearby clusters.  That is, the phase space distribution of galaxies in
redshift and projected radius contains a dense envelope of galaxies
that is well separated from foreground and background galaxies.
Surprisingly, the infall patterns or ``caustics'' have significantly
higher contrast in the CAIRNS observations than in the simulations of
DG and D99.  The CAIRNS clusters are fairly massive clusters and
generally have relatively little surrounding large-scale structure
\citep[but see][]{rines01b,rines02}.  One might suspect that the
presence of infall patterns is limited to massive, isolated clusters.
However, other investigators have found infall patterns around the
Fornax Cluster \citep{drink}, the Shapley Supercluster
\citep{rqcm}, an ensemble cluster comprised of poor clusters in the
Two Degree Field Galaxy Redshift Survey \citep{bg03}, and even the
galaxy group associated with NGC 5846 \citep{mahdavi05}.

CAIRNS showed that caustic masses of clusters agree well with mass
estimates from both X-ray observations and Jeans' analysis at small
radii \citep{cairnsi}.  \citet{lokas03} confirm that the mass of Coma
estimated from higher moments of the velocity distribution agrees well
with the caustic mass estimate \citep{gdk99}.  Recently,
\citet{diaferio05} showed that caustic masses agree with weak
lensing masses in three clusters at moderate redshift.

Although the presence of caustics in all nine CAIRNS clusters suggests
that they are ubiquitous, some readers may wonder how robust the
caustic technique is.  To that end, we use the Fourth Data Release
(DR4) of the Sloan Digital Sky Survey (SDSS) to test the ubiquity of
infall patterns around galaxy clusters.  We match four X-ray cluster
catalogs derived from the ROSAT All-Sky Survey \citep[RASS;][]{rass}
to the spectroscopic area covered in DR4 and analyze the resulting
sample of 72 clusters, which we refer to as the CIRS (Cluster Infall
Regions in SDSS) sample.  Our procedure is similar to the RASS-SDSS
\citep{popesso04,popesso05} analysis of clusters in DR2, but we
concentrate on the outskirts of clusters and on nearby systems with
large redshift samples.  \citet{miller05} construct a cluster catalog
by locating overdensities in position and color within SDSS DR2.

CAIRNS also provides an important zero-redshift benchmark for
comparison with more distant systems \citep[e.g.,][]{ellingson01}.  In
particular, the caustic technique is a membership algorithm. The
resulting samples should therefore provide clean measures of cluster
scaling relations.  The CNOC1 project assembled an ensemble cluster
from X-ray selected clusters at moderate redshifts. The CNOC1 ensemble
cluster samples galaxies up to $\sim$2 virial radii \citep[see][and
references therein]{cye97,ellingson01}.  The caustic pattern is easily
visible in the ensemble cluster, but \citet{cye97} apply only Jeans
analysis to the cluster to determine an average mass profile.
\citet{bg03} analyzed cluster redshifts from the 2dF 100,000 redshift
data release. They stacked 43 poor clusters to produce an ensemble
cluster containing 1345 galaxies within 2 virial radii and analyzed
the properties of the ensemble cluster with both Jeans analysis and
the caustic technique.  \citet{bg03} find good agreement between the
two techniques; the caustic mass profile beyond the virial radius
agrees well with an extrapolation of the Jeans mass profile.  In
contrast to these studies, the CAIRNS clusters are sufficiently well
sampled to apply the caustic technique to the individual clusters.
Similarly, SDSS provides sufficiently deep and dense samples that we
can study the infall regions of individual clusters, thus constraining
cluster-to-cluster variations and avoiding uncertainties in stacking
procedures.

The primary advantage of the CIRS sample is the large number of
clusters.  Besides confirming the CAIRNS results, CIRS provides a
sufficiently large sample of clusters to enable much more general
constraints on the properties of clusters and their dark matter
profiles.  CIRS also covers a wider range of cluster masses, X-ray
luminosities, and environments.  The larger number of clusters enables
new comparisons of cluster properties (e.g., the shapes of dark matter
profiles on large scales) with simulations.

We describe the data and the cluster sample in $\S$ 2.  In $\S$ 3, we
review the caustic technique and use it to estimate the cluster mass
profiles, discuss cluster scaling relations, and compare the caustic
mass profiles to simple parametric models.  We discuss some individual
clusters in $\S 4$. We compare the caustic mass profiles to X-ray and
virial mass estimators in $\S$ 5.  We compute the velocity dispersion
profiles in $\S$ 6.  We summarize our results in $\S 7$.  We assume
$H_0 = 100 h~\kms, \Omega _m = 0.3, \Omega _\Lambda = 0.7$ throughout.

\section{The CIRS Cluster Sample}

\subsection{Sloan Digital Sky Survey \label{sdssdesc}}

The Sloan Digital Sky Survey \citep[SDSS,][]{sdss} is a wide-area
photometric and spectroscopic survey at high Galactic latitudes.  The
Fourth Data Release (DR4) of SDSS includes 6670 square degrees of
imaging data and 4783 square degrees of spectroscopic data
\citep{dr4}.

The spectroscopic limit of the main galaxy sample of SDSS is $r$=17.77
after correcting for Galactic extinction \citep{strauss02}.  CAIRNS
sampled all galaxies brighter than about $M^*+1$ and often fainter.
Figure 7 of \citet{cairnsha} shows the infall patterns for galaxies
brighter than $M_{K_s}^*+1$.  These patterns are readily apparent, but
less well-defined than the full CAIRNS samples \citep{cairnsi}.
Assuming the luminosity function of \citet{blanton03}, the
spectroscopic limit of SDSS corresponds to $M_{^{0.1}r}^*+1$ at
$z=0.092$.  Thus, we expect that infall patterns of clusters with
masses similar to CAIRNS clusters should be apparent to
$z\lesssim$0.1, though not much further.

Note that the SDSS Main Galaxy Survey is $\sim$85-90\% complete to the
spectroscopic limit.  The survey has $\approx$7\% incompleteness due
to fiber collisions \citep{strauss02}, which are likely more common in
dense cluster fields.  Because the target selected in a fiber
collision is determined randomly, this incompleteness can
theoretically be corrected for in later analysis.  From a comparison
of SDSS with the Millennium Galaxy Catalogue,
\citet{2004MNRAS.349..576C} conclude that there is an additional
incompleteness of $\sim$7\% due to galaxies misclassified as stars or
otherwise missed by the SDSS photometric pipeline.  For our purposes,
the incompleteness is not important provided sufficient numbers of
cluster galaxies do have spectra.

\subsection{X-ray Cluster Surveys \label{xcs}}

Because SDSS surveys primarily low-redshift galaxies, the best sampled
clusters are both nearby and massive.  We therefore search X-ray
cluster catalogs derived from the ROSAT All-Sky Survey for clusters in
DR4.  RASS \citep{rass} is a shallow survey but it is sufficiently
deep to include nearby, massive clusters.  RASS covers virtually the
entire sky and is thus the most complete X-ray cluster survey for
nearby clusters.  RASS was conducted with the PSPC (Position Sensitive
Proportional Counter) instrument.

Published cluster catalogs derived from the RASS include the X-ray
Brightest Abell Cluster Survey (XBACS), the Bright Cluster Survey
(BCS), the NOrthern ROSAT All-Sky galaxy cluster survey (NORAS), and
the ROSAT-ESO flux limited X-ray  galaxy cluster survey
(REFLEX).  XBACS is ``an essentially complete, all-sky, X-ray
flux-limited sample of 242 Abell clusters'' from RASS I \citep{xbacs},
the first processing of the RASS.  BCS is the Bright Cluster Survey
\citep{bcs} and eBCS is an extension of this survey to a smaller
limiting flux \citep{2000MNRAS.318..333E}.  XBACs and BCS/eBCS were
constructed first by searching for X-ray sources around known optical
clusters from the Abell and Zwicky catalogs; they then use a Voronoi
Tesselation and Percolation (VTP) algorithm to identify extended X-ray
sources (from a master catalog of X-ray point sources) not associated
with optical clusters.  PSPC count rates are converted to total fluxes
(in the 0.1-2.4 keV energy band) using King models to account for
missing flux at large radii and corrected for Galactic absorption and
$K$ corrected using either temperatures from \citet{david93} or a
correction from the $L_X-T$ relation from an iterative process.
BCS/eBCS covers the northern hemisphere at high Galactic latitudes
($\delta \geq$0$^\circ$ and $|b|>$20$^\circ$).  The flux limits of BCS
and eBCS are 4.4 and 2.8$\times 10^{-12}$erg s$^{-1}$ cm$^{-2}$
respectively.  \citet{bcs} estimate the completeness to these limits
is approximately 90\% and 80\%, where the missing clusters are
extended sources not included in the master catalog of X-ray point
sources.

NORAS \citep{2000ApJS..129..435B} also covers the northern hemisphere
at high Galactic latitudes ($\delta \geq$0$^\circ$ and
$|b|>$20$^\circ$).  The initial source catalog consists of extended
objects detected in RASS I with their properties determined from RASS
II \citep[the second processing of RASS, ][]{rass} data.  The fluxes
are computed using a method called growth curve analysis and are
corrected for missing flux by fitting the emission profile to a
$\beta$ model \citep{cff} and extrapolating to 12 core radii.  The
NORAS fluxes agree well with BCS/eBCS fluxes for clusters in common.
Both surveys are expected to be incomplete.  Comparing NORAS and
REFLEX, \citet{noras} estimate that the completeness of NORAS is about
50\%.  In a subset of NORAS (an area of sky between right ascensions
9$^h$ and 14$^h$ and declinations $\delta$$>$0), the completeness
increases to 82\% with the addition of several additional clusters
found in a thorough search of Abell cluster positions and extended
sources in the RASS II catalog \citet{noras}.  We include these
clusters in our sample.

REFLEX \citep{2001A&A...369..826B,2004A&A...425..367B} is analogous to
NORAS but for the southern hemisphere.  Both NORAS and REFLEX extend
to smaller X-ray fluxes than XBACS.  The catalog construction is
different for each survey: XBACs and BCS/eBCS initially search for
extended X-ray sources around known optical clusters; BCS/eBCS
includes additional X-ray selected clusters selected from a master
catalog of point sources.  NORAS is selected from extended X-ray
sources with no explicit optical selection.  Finally, REFLEX matches
X-ray flux overdensities with galaxy overdensities and is therefore
selected jointly in X-ray and optical wavelengths.  The REFLEX fluxes
are calculated in the same manner as NORAS.

When multiple X-ray fluxes are available for a cluster, we use the
most recently published value.  The order of preferences is therefore:
REFLEX, NORAS, BCS/eBCS, XBACs.  We define a flux-limited and
redshift-limited sample of clusters with the criteria $f_X\geq 3\times
10^{-12}$erg s$^{-1}$cm$^{-2}$ (0.1-2.4 keV) and $z\leq$0.10.  Note
that variations in the method of determining flux in different
catalogs may affect the precise flux limit.  The only cluster with an
XBACs flux is A1750b, the southern component of A1750.  We follow
\citet{2004A&A...425..367B} and treat A1750 as a single source rather
than two separate sources as in XBACs.  Similarly, the galaxy NGC5813
is bound to the NGC5846 group \citep{mahdavi05}.  Because the dynamics
of NGC5813 are dominated by the NGC5846 system, we eliminate NGC5813
from the sample.  We inspect the redshift data around each cluster to
confirm the cluster redshift and find that A2064 has an incorrect
redshift (and X-ray luminosity) listed in NORAS.  The correct redshift
is 0.0738 instead of 0.1076.  We correct the X-ray luminosity
accordingly.  The redshift of A2149 is listed as $z$=0.1068 in NORAS
and $z$=0.0675 in eBCS.  Because the X-ray peak of the RASS image lies
near an apparent BCG at the latter redshift (see
$\S$\ref{individual}), we adopt the eBCS value and adjust the X-ray
luminosity accordingly.  Our final flux and redshift limited sample
contains \ncirs~clusters within the SDSS DR4 spectroscopic footprint.
We will refer to this sample as the CIRS (Cluster Infall Regions in
SDSS) clusters hereafter.  The completeness of the CIRS sample is
limited by the completeness of the underlying cluster catalogs.
However, by combining clusters from the various catalogs we should be
more complete than any individual catalog.  For our purposes, this
modest potential incompleteness is not important.  The clusters are an
unbiased sample: the selection of the CIRS sample is based purely on
X-ray flux and the footprint of the SDSS DR4 spectroscopic survey.

\section{Results}

\subsection{Ubiquity of Infall Patterns around Clusters}

We first test for the ubiquity of caustic or infall patterns around
X-ray clusters.  Analogous to CAIRNS, we plotted radius-redshift
diagrams for all clusters in the X-ray cluster catalogs covered by DR4
with $z<$0.20.  Clusters with $z>$0.10 have few spectroscopically
confirmed members in DR4 (see $\S$\ref{sdssdesc}).  The caustic
pattern is a superposition of the ``finger-of-God'' elongation in
redshift space of the virialized galaxies in the cluster center and
the flattening due to infall at large radii.  The resulting infall
pattern is a characteristic trumpet shape in phase space
\citep[][D99]{kais87}.  Substructure in the infall region smears out
the sharp enhancements in phase space predicted by simple spherical
infall, so the infall pattern should be apparent as a dense envelope
of galaxies in phase space with well-defined edges (D99).  The
detailed appearance of the redshift-radius diagrams depends on the
line-of-sight of the observer, so the contrast between the dense
infall envelope and foreground and background galaxies is subject to
projection effects (D99).  We assign a ``by-eye'' classification of
each cluster's infall pattern: ``clean'' for clusters with few
background and foreground galaxies, ``intermediate'' for clusters with
apparent infall patterns but significant contamination from either
related large-scale structure or foreground and background objects,
and ``none'' for clusters with no apparent infall pattern.  Below, we
demonstrate that the caustic technique can be applied successfully to
clusters classified here as ``intermediate'' or ``none''; our
classification scheme is thus fairly conservative.  We use this
classification scheme {\it only} to show the dependence of the infall
pattern appearance on cluster mass and the sampling depth.

\begin{figure}
\figurenum{1}
\plotone{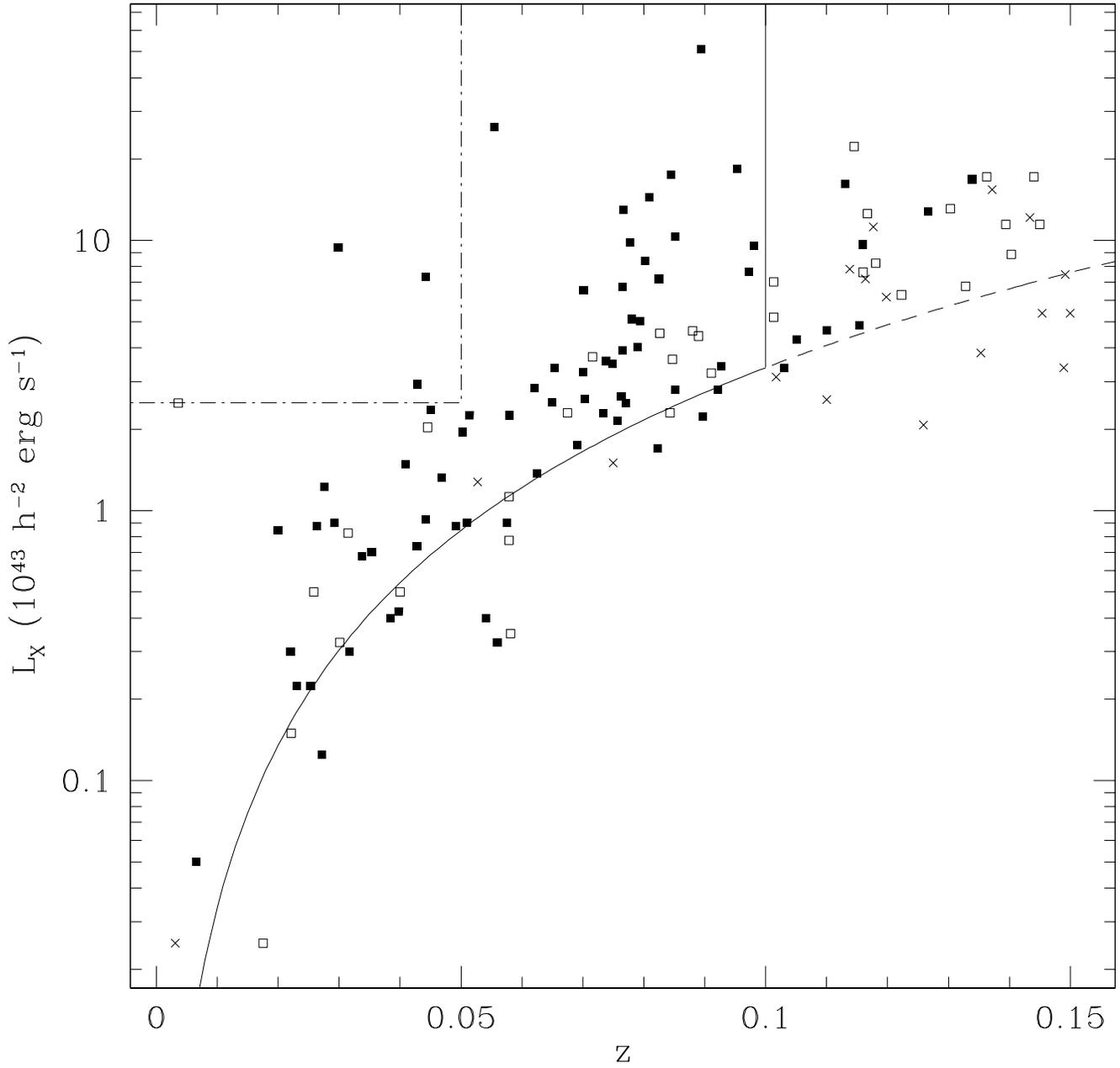}
\caption{\label{cirslxz} Redshift versus X-ray luminosity (0.1-2.4 keV) 
for X-ray clusters from XBACS, BCS/eBCS, NORAS, and REFLEX contained
in the SDSS DR4 spectroscopic survey region.  Filled squares, open
squares and crosses indicate clusters with ``clean'' infall patterns,
``intermediate'' infall patterns, and no obvious infall pattern respectively.
The X-ray cluster catalogs are complete to approximately
$f_X>$3$\times$10$^{-12}$erg s$^{-1}$ (curved line). The solid line
shows the flux and redshift limits of the CIRS cluster sample.  The
dash-dotted line shows the redshift and luminosity limits of CAIRNS. }
\end{figure}

Figure \ref{cirslxz} shows the dependence of this subjective
classification on X-ray luminosity and redshift.  As expected from the
depth of SDSS ($\S \ref{sdssdesc}$), there are few ``clean'' infall
patterns where SDSS samples shallower than $M_*+1$.  We therefore
define a ``complete'' sample selected by X-ray flux $f_X\geq 3\times
10^{-12}$erg s$^{-1}$cm$^{-2}$ (0.1-2.4 keV) and redshift
($z\leq$0.10).  The CIRS sample contains \ncirs ~clusters; the vast
majority of this sample (56/\ncirs~or 78\%) contain ``clean''
infall patterns and only three (4\%) show no obvious infall pattern.
Of the three clusters in the latter category, two are embedded in
larger structures (NGC4636 in the outskirts of the Virgo cluster and
A2067 in the outskirts of A2061, part of the CorBor supercluster) and
one (A1291) is a possible merging cluster with two nearly concentric
components separated by $\sim$2000~$\kms$ (see $\S$\ref{individual} for
details).  In $\S$3.2 we are able to identify caustics in 100\% of the
CIRS clusters.  Figure \ref{cirslxz} also shows the CAIRNS X-ray
luminosity and redshift limits.  The three CIRS clusters meeting this
criteria were studied by CAIRNS (A119, A168, and A2199).  Figure
\ref{cirslxz} demonstrates the expanded parameter space covered by the
CIRS sample.  Table \ref{sample} lists the clusters in the CIRS
sample, their X-ray positions, luminosities, and temperatures (when
available), their central redshifts and velocity dispersions (see
below), and the projected radius $R_{comp}$ within which the SDSS DR4
spectroscopic survey provides complete spatial coverage.  For several
clusters, the caustic pattern disappears beyond $R_{comp}$ because of
this edge effect.
%The only
%cluster with $L_X>$2$\times10^{43} h^{-2}$erg s$^{-1}$ that does {\it not} display a clear infall
%pattern is Virgo; because we are inside Virgo, the infall pattern
%looks different than distant clusters.    
Figures \ref{allcirs1}-\ref{allcirs6} show the radius-redshift
diagrams for the CIRS sample.  Vertical lines in some panels indicate
the radius within which the DR4 spectroscopic footprint provides
complete coverage of the infall region; the absence of a caustic
pattern in some clusters beyond this limit is due to incomplete
spatial coverage.
%Figure \ref{allcirs6} shows the infall
%patterns for secondary components in A1035 and A1291 (discussed in
%$\S$\ref{individual}). 
%and in A1383, an X-ray cluster in the outskirts
%of A1377.  A1383 has a measured flux of 1.4$\times 10^{-12}$erg
%s$^{-1}$cm$^{-2}$ \citep{2003AJ....126.2740L}, although they note that
%their fluxes are measured in a smaller aperture than BCS/eBCS and are
%therefore about a factor of 1.8 smaller than BCS/eBCS fluxes.

Another factor which determines the presence or absence of a ``clean''
infall pattern is the surrounding large-scale structure.  For example,
the redshift-radius diagrams of groups within the infall regions of
more massive clusters reflect the kinematics of the cluster's infall
region \citep[e.g., A2199, see][]{rines02}.  The CIRS sample contains
several of these systems: NGC6107 and A2197 in the infall region of
A2199, NGC4636 in the infall region of Virgo, NGC5813 in the infall
region of NGC5846, and A2067 in the infall region of A2061.  A1173 and
A1190 are close and likely bound.  Two clusters, A1035 and A1291, show
evidence of two infall patterns in the projected radius-redshift
diagrams. We label the two components A and B with A indicating the
lower redshift component.  The caustic algorithm finds A1035B and
A1291A to be the larger components; we use these components
when compiling results for the CIRS sample.  We discuss these systems
in more detail in $\S \ref{individual}$.

\subsection{Caustics and Mass Profiles}

Similar to CAIRNS, the CIRS clusters show infall patterns much more
well defined than those of the simulations of D99.
Because infall patterns are better defined in a low-density universe
(D99), a possible explanation for this discrepancy is
that the real universe has a smaller matter density than the
simulations.  Another possibility is that the semi-analytic galaxy
formation recipes used in D99 are inaccurate around
massive clusters.

We briefly review the method DG and D99 developed to estimate the mass
profile of a galaxy cluster by identifying caustics in redshift space.
The method assumes that clusters form in a hierarchical process.
Application of the method requires only galaxy redshifts and sky
coordinates.  Toy models of simple spherical infall onto clusters
produce sharp enhancements in the phase space density around the
system.  These enhancements, known as caustics, appear as a trumpet
shape in scatter plots of redshift versus projected clustercentric
radius \citep{kais87,rg89}.  DG and D99 show that random
motions smooth out the sharp pattern expected from simple spherical
infall into a dense envelope in the redshift-radius diagram \citep[see
also][]{vh98}.  The edges of this envelope can be interpreted as the
escape velocity as a function of radius.  Galaxies outside the
caustics are also outside the turnaround radius.  The caustic
technique provides a well-defined boundary between the infall region
and interlopers; one may think of the technique as a method for
defining membership that gives the cluster mass profile as a
byproduct.

The amplitude $\mathcal{A} \mathnormal{(r)}$ of the caustics is half of
the distance between the upper and lower caustics in redshift
space. Assuming spherical symmetry, $\mathcal{A} \mathnormal(r)$ is
related to the cluster gravitational potential $\phi (r)$ by
\beqn
\mathcal{A} \mathnormal {^2 (r) = -2 \phi (r)\frac{1-\beta (r)}{3-2\beta (r)}}
\eeqn
where $\beta (r) = 1- [\sigma_\theta^2(r)/\sigma_r^2(r)]$ is the
velocity anisotropy parameter and $\sigma_\theta$ and $\sigma_r$ are
the tangential and radial velocity dispersions respectively. DG show
that the mass of a spherical shell of radii [$r_0,r$] within the
infall region is the integral of the square of the amplitude
$\mathcal{A}
\mathnormal{(r)}$
\beqn
%\label{mprofeqn}
GM(<r)-GM(<r_0) = F_\beta \int_{r_0}^r \mathcal{A} \mathnormal{^2(x)dx}
\eeqn
where $F_\beta \approx 0.5$ is a filling factor with a numerical value
estimated from simulations. Variations in $F_\beta$ lead to some
systematic uncertainty in the derived mass profile (see D99 for a more
detailed discussion).  Note that the variations in $F_\beta$ are
included in the estimates of the systematic uncertainties in the
caustic technique.

Operationally, we identify the caustics as curves which delineate a
significant decrease in the phase space density of galaxies in the
projected radius-redshift diagram.  For a spherically symmetric
system, taking an azimuthal average amplifies the signal of the
caustics in redshift space and smooths over small-scale substructures.
We isolate the clusters initially by studying only galaxies within
$R_p\leq$10$\Mpc$ and $\pm$5000$\kms$ of the nominal cluster centers
from the X-ray catalogs.  We perform a hierarchical structure analysis
to locate the centroid of the largest system in each volume (see
Appendix A of D99).  This analysis consists of creating a binary tree
based on estimated binding energies, identifying the largest cluster
in the field, and determining its center from the two-dimensional
distribution of celestial coordinates determined with adaptive kernel
smoothing.  This analysis sometimes finds the center of another system
in the field.  In these cases, limiting the galaxies to a smaller
radial and/or redshift range enables the algorithm to center on the
desired cluster.  Table \ref{centerproblems} lists these restrictions.

We adaptively smooth the azimuthally averaged phase space diagram (the
ensemble of redshift, radius data points) and the algorithm chooses a
threshold in phase space density as the edge of the caustic envelope.
The upper and lower caustics at a given radius are the redshifts at
which this threshold density is exceeded when approaching the central
redshift from the ``top'' and ``bottom'' respectively of the
redshift-radius diagram.  Because the caustics of a spherical system
are symmetric, we adopt the smaller of the upper and lower caustics as
the caustic amplitude $\mathcal{A}\mathnormal{(r)}$ at that radius.
This procedure reduces the systematic uncertainties introduced by
interlopers, which generally lead to an overestimate of the caustic
amplitude.  The threshold is defined by the algorithm to minimize the
quantity $|<$$v_{esc}^2$$>_R - 4 <$$v^2$$>_R|$, where R is a virial-like
radius (see D99 for details).  \citet{rines02} explore the effects of
altering some of these assumptions and find that the differences are
generally smaller than the estimated uncertainties.

D99 described this method in detail and showed that, when applied to
simulated clusters containing galaxies modelled with semi-analytic
techniques, it recovers the actual mass profiles to radii of
5-10$~\Mpc$ from the cluster centers.  D99 gives a prescription for
estimating the uncertainties in the caustic mass profiles.  The
uncertainties estimated using this prescription reproduce the actual
differences between the caustic mass profiles and the true mass
profiles of the simulated clusters including systematic effects such
as departures from spherical symmetry.  The uncertainties in the
caustic mass profiles of observed clusters may be smaller than the
50\% uncertainties in the simulations \citep{cairnsi}.  This
difference is due in part to the large number of redshifts in the CIRS
redshift catalogs relative to the simulated catalogs.  Furthermore,
the caustics are generally more cleanly defined in the data than in
the simulations.  Clearly, more simulations which better reproduce the
appearance of observed caustics and/or include fainter galaxies would
be useful in determining the limits of the systematic uncertainties in
the caustic technique.  We are currently conducting an analysis for
the hydrodynamical simulations of \citet{borgani04} with these goals
(A.~Diaferio et al.~2006, in preparation).

We calculate the shapes of the caustics with the technique described
in D99 using a smoothing parameter of $q$=25.  The smoothing parameter
$q$ is the scaling between the velocity smoothing and the radial
smoothing in the adaptive kernel estimate of the underlying phase
space distribution (e.g., a particle which has a smoothing window in
the radial direction of 0.04$h^{-1}$Mpc will have a smoothing window
in the velocity direction of 100 $\kms$ for $q$=25 and 40 $\kms$ for
$q$=10).  Previous investigations show that the mass profiles are
insensitive to changes of a factor of 2 in the smoothing parameter
\citep{gdk99,rines2000,rines02}.  

\begin{figure}
\figurenum{2}
\plotone{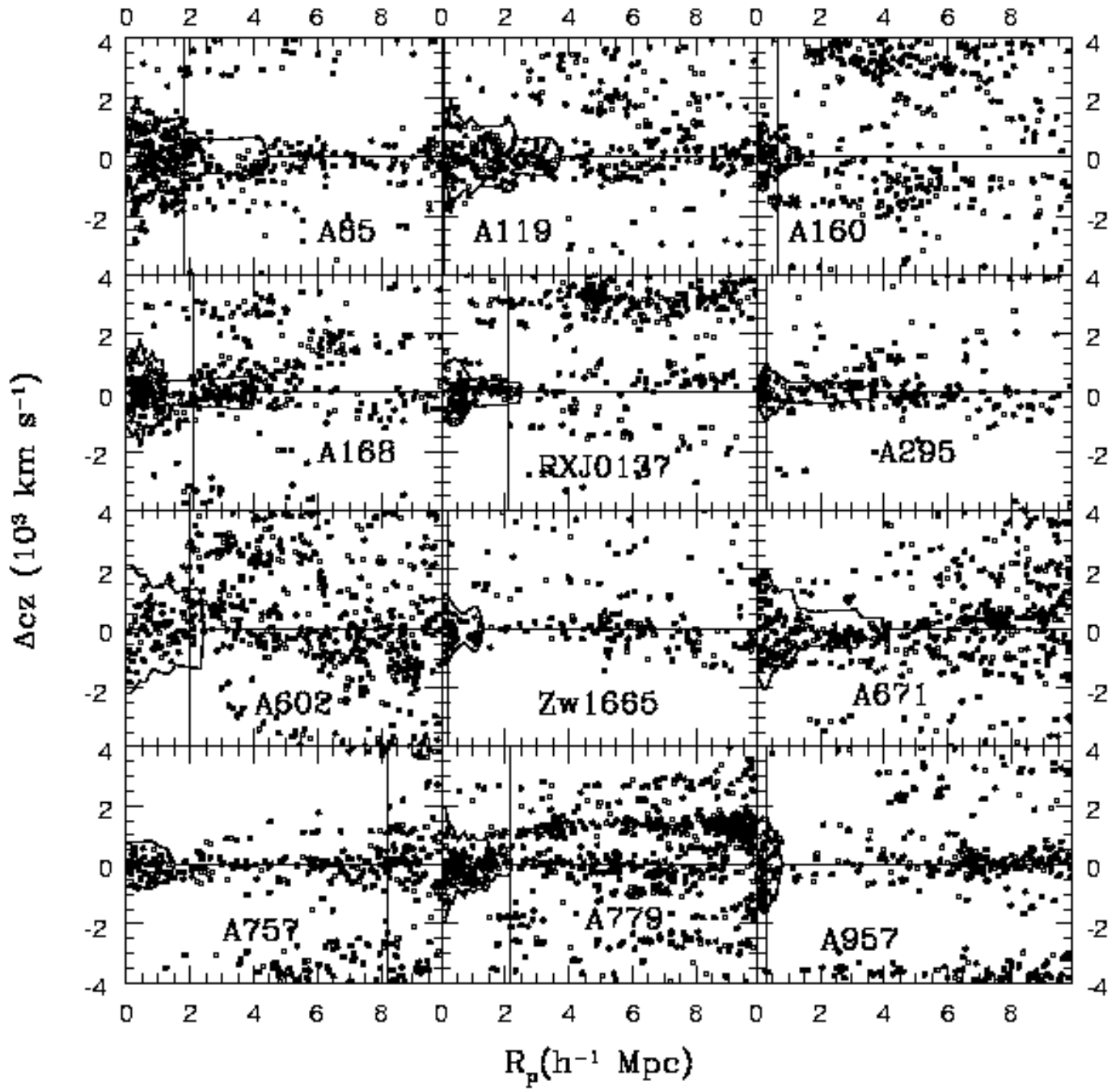}
\caption{\label{allcirs1} Redshift versus radius for SDSS galaxies around
the first twelve X-ray clusters in the CIRS sample.  The caustic
pattern is evident as the trumpet-shaped regions with high density.
The solid lines indicate our estimate of the location of the caustics
in each cluster.  
%The errorbars are 1$\sigma$ uncertainties and are
%shown only on one side of each caustic for clarity.  
Vertical lines in each panel
indicate the radius where the spatial coverage of the SDSS DR4
spectroscopic survey is no longer complete.  Figures
\ref{allcirs2}-\ref{allcirs6} show similar plots for the rest of the
sample.}
\end{figure}

\begin{figure}
\figurenum{3}
\plotone{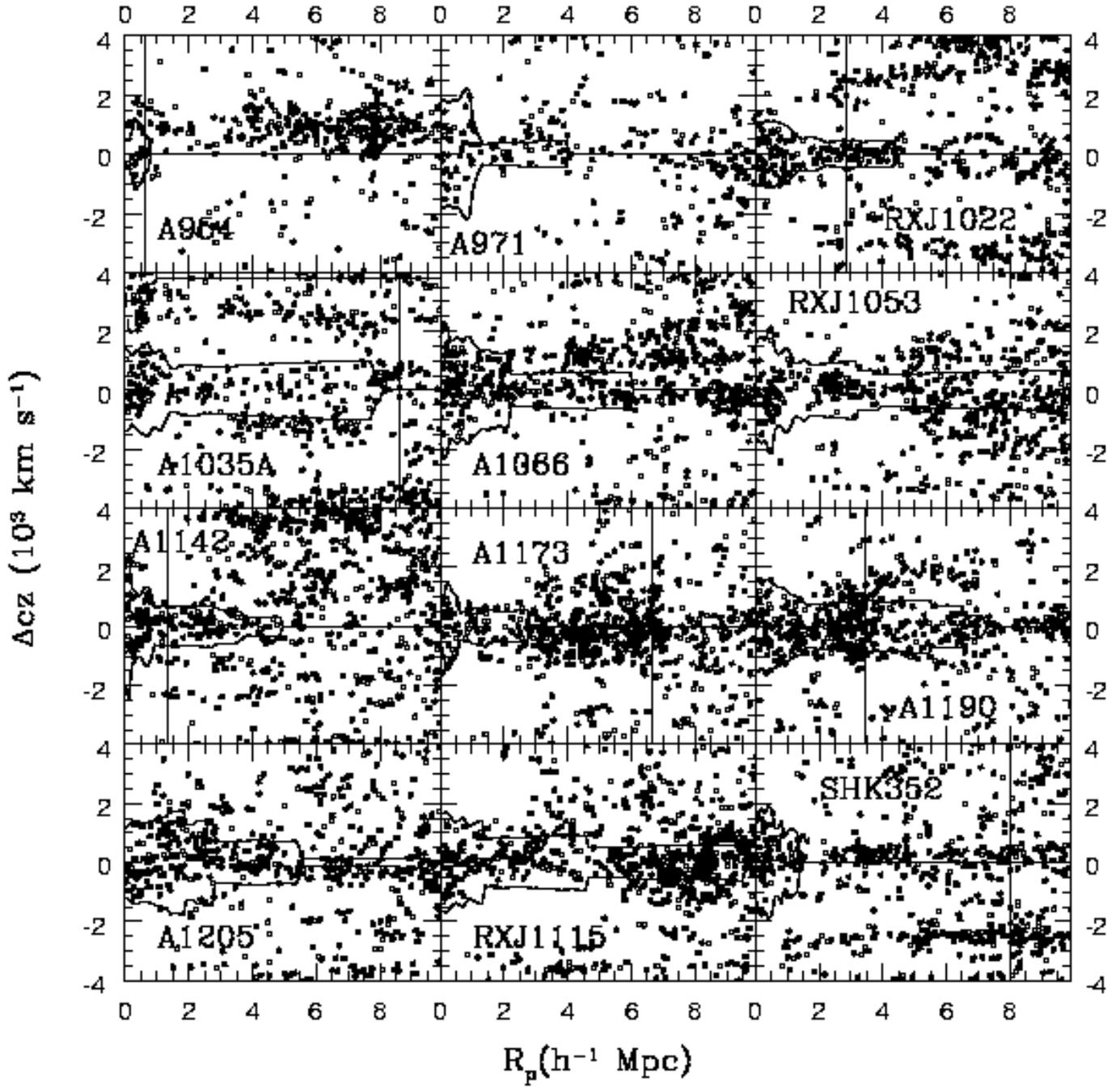}
\caption{\label{allcirs2} See Figure \ref{allcirs1}.  The dashed lines in the diagram for A1035A indicate the infall pattern of a higher-redshift component.}  
\end{figure}

\begin{figure}
\figurenum{4}
\plotone{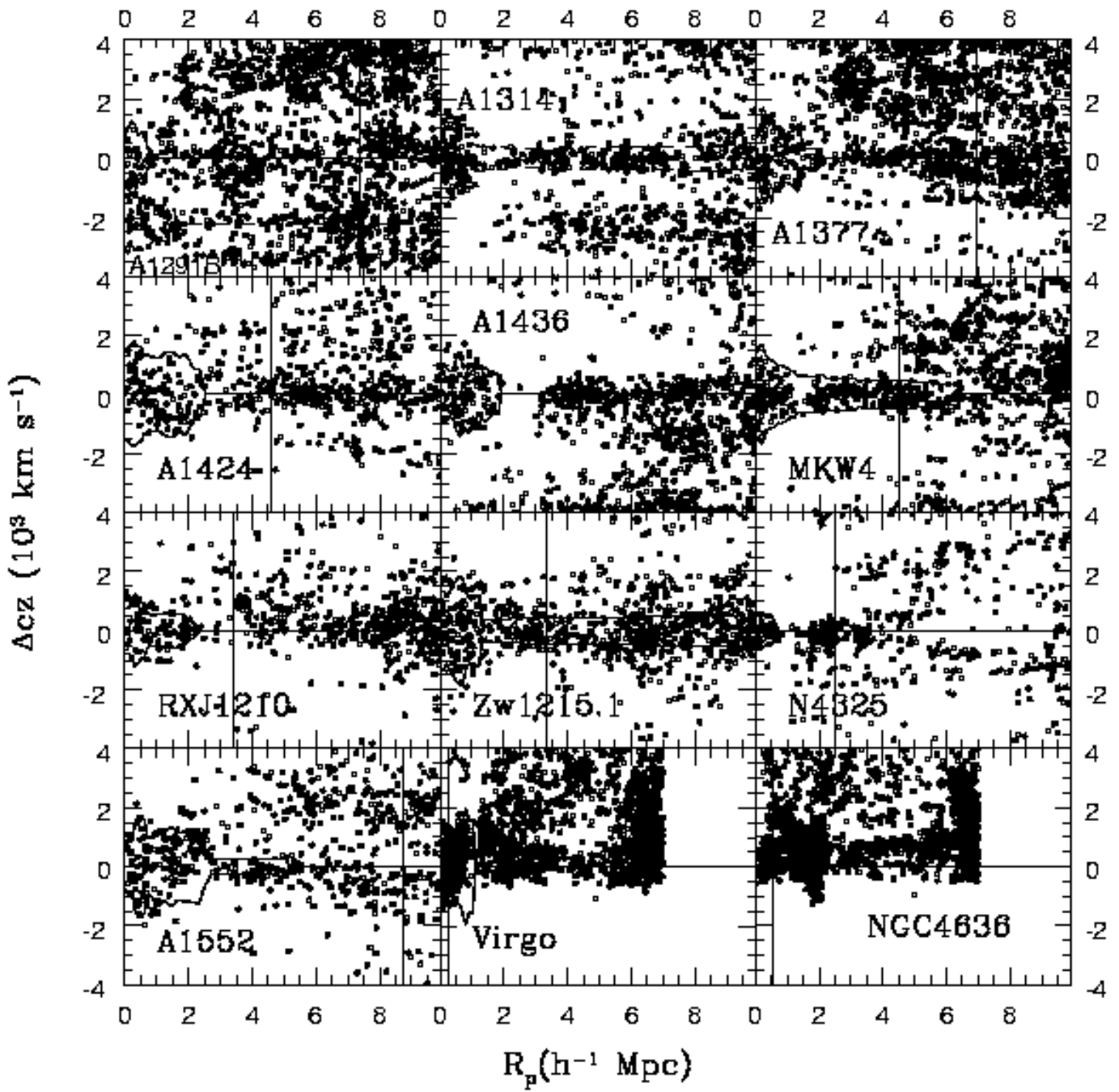}
\caption{\label{allcirs3} See Figure \ref{allcirs1}.  The dashed lines in the diagram for A1291B indicate the infall pattern of a lower-redshift component.}  
\end{figure}

\begin{figure}
\figurenum{5}
\plotone{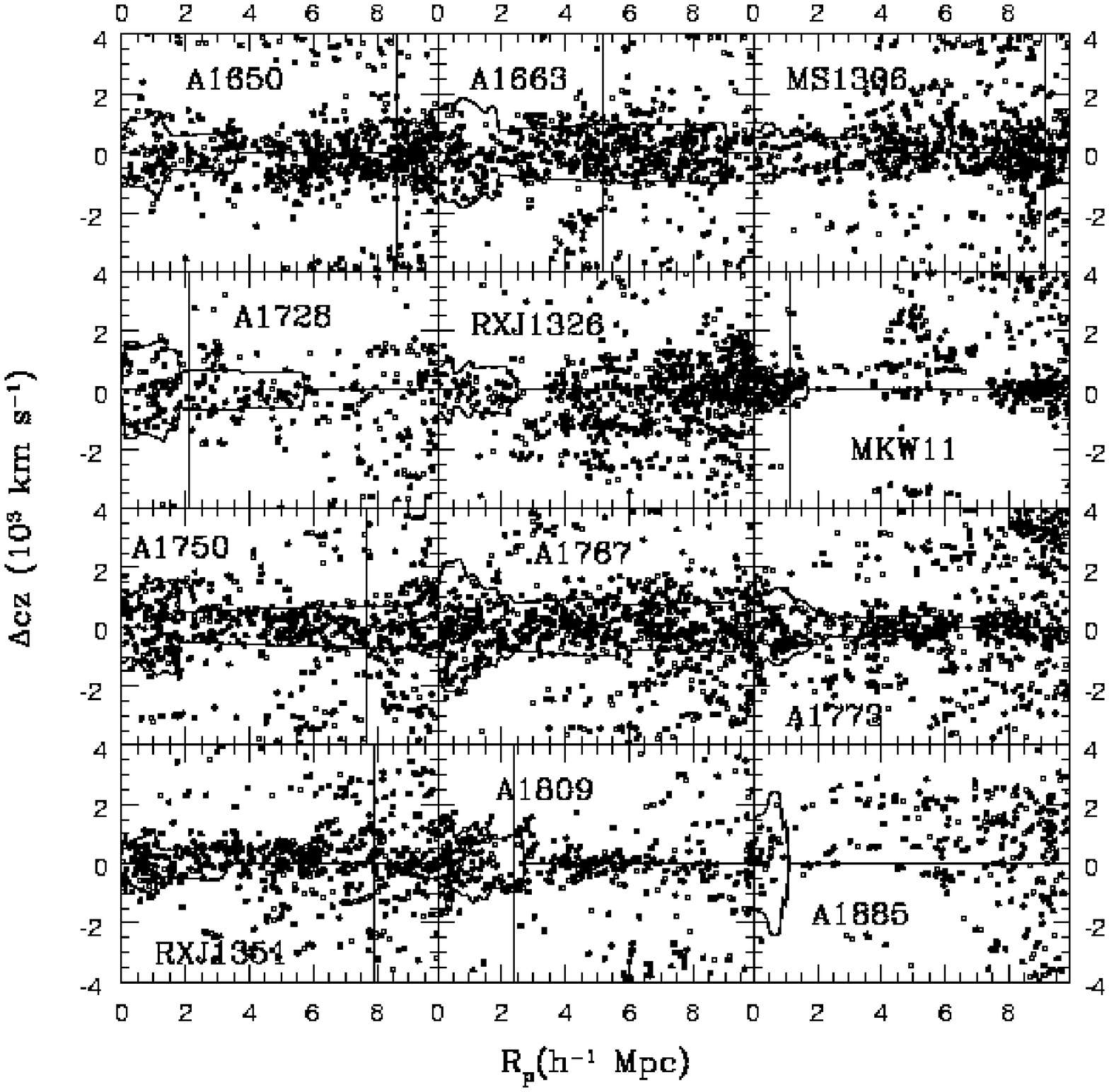}
\caption{\label{allcirs4} See Figure \ref{allcirs1}.}  
\end{figure}

\begin{figure}
\figurenum{6}
\plotone{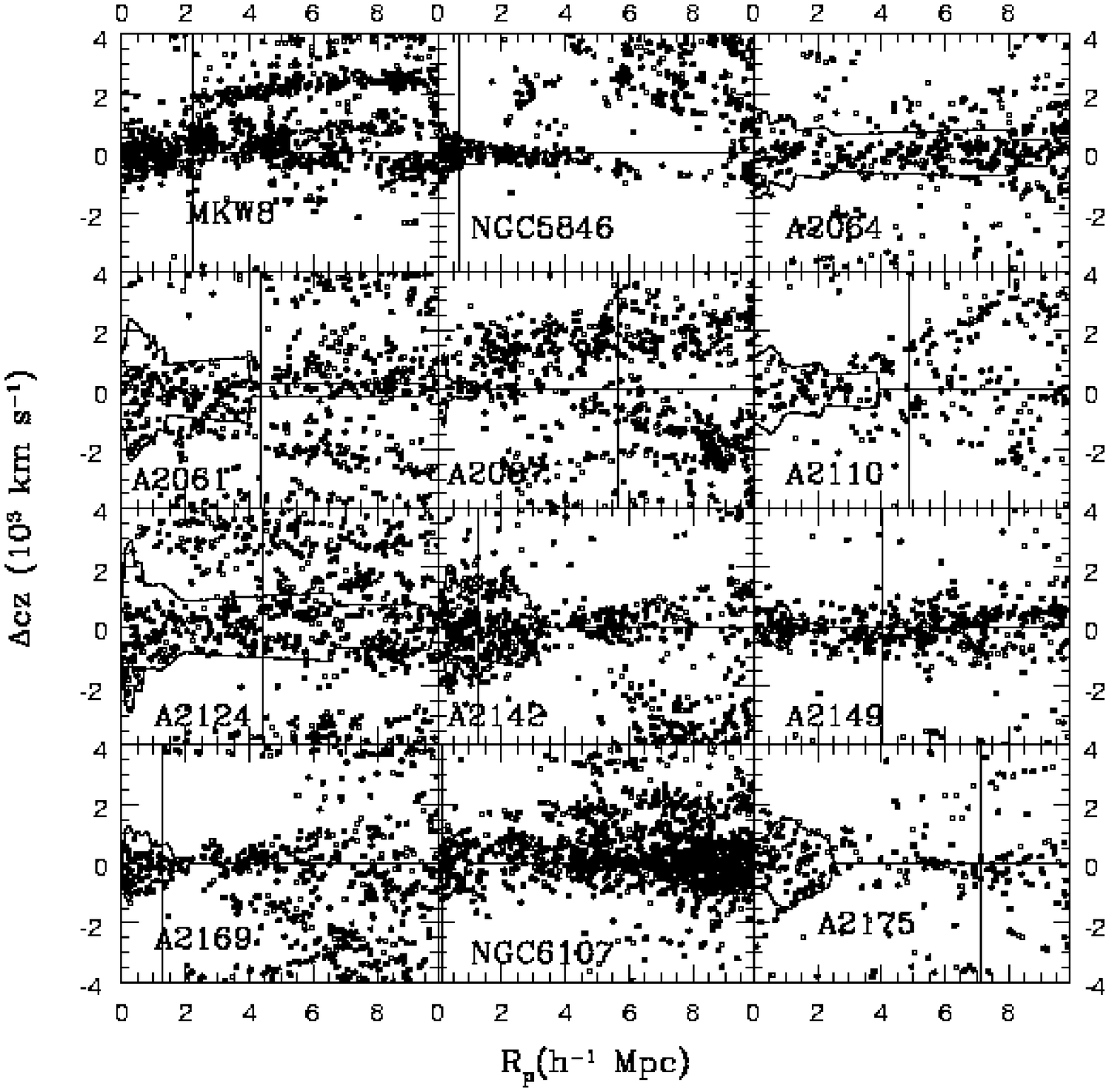}
\caption{\label{allcirs5} See Figure \ref{allcirs1}.}  
\end{figure}

\begin{figure}
\figurenum{7}
\plotone{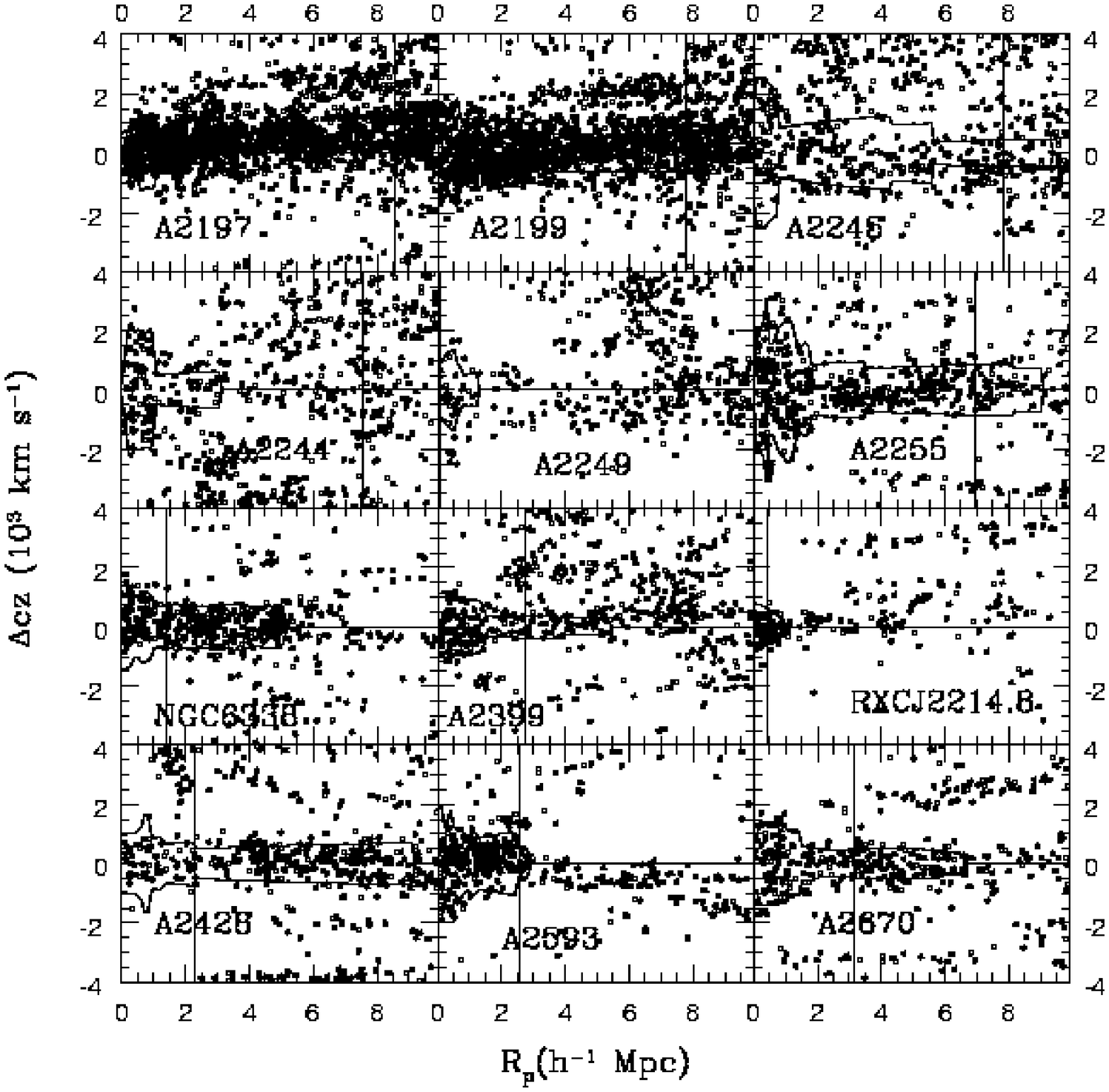}
\caption{\label{allcirs6} See Figure \ref{allcirs1}.}  
\end{figure}

Table \ref{centers} lists the hierarchical centers.  These centers
generally agree with the X-ray positions (Table \ref{centers}) with a
median difference of 109$~\kpc$ and with the redshift centers from the
X-ray catalogs with a median difference of -17 $\kms$.  The
hierarchical centers disagree by more than 500$\kpc$ in four clusters
and by more than 1000 $\kms$ for seven clusters.  The hierarchical
redshift centers are more reliable than those in the X-ray cluster
catalogs because the former are based on much larger redshift samples.
We discuss some of the individual cases in $\S$\ref{individual}.
Figures \ref{allcirs1}-\ref{allcirs6} show the caustics and Figures
\ref{allcirsm1}-\ref{allcirsm6} show the associated mass profiles.
Note that the caustics extend to different radii for different
clusters.  D99 show that the appearance of the caustics depends
strongly on the line of sight; projection effects can therefore
account for most of the differences in profile shape in Figures
\ref{allcirs1}-\ref{allcirs6} without invoking non-homology among
clusters.  The uncertainties in the caustic mass profiles are
estimated with the prescription of D99 for Coma-size clusters.  Under
this prescription, more densely sampled clusters and those with higher
contrast between the caustics and the background have smaller
uncertainties than sparsely sampled clusters or those with poorer
contrast.  Thus, the uncertainties in the caustic mass profiles for
some CIRS clusters (those with smaller masses or unusually empty
backgrounds) computed with this prescription may underestimate the
total systematic uncertainties.

We use the caustics to determine cluster membership.  Here,
the term ``cluster member'' refers to galaxies both in the virial
region and in the infall region.  Figures
\ref{allcirs1}-\ref{allcirs6} show that the caustics effectively
separate cluster members from background and foreground galaxies,
although some interlopers may lie within the caustics. This clean
separation affirms our adoption of velocity dispersions calculated
from cluster members as defined by the caustics ($\S$3.2).

We find a measureable signal for the caustic profile for all 72 X-ray
clusters in the CIRS sample (that is, all show an identifiable cluster
of galaxies with a surrounding infall pattern).  This amazing success
rate demonstrates the power and ubiquity of the caustic technique in
identifying the galaxies associated with clusters and their infall
regions.

In the simulations of D99, the degree of definition of the caustics
depends on the underlying cosmology; caustics are better defined in a
low-density universe than a flat, matter-dominated universe.
Surprisingly, the contrast of the phase space density between regions
inside and outside the caustics is much stronger in the data than in
both the $\tau$CDM and $\Lambda$CDM simulated clusters in D99.  The
difference may arise from the cosmological model used or the
semi-analytic techniques for defining galaxy formation and evolution
in the simulations.  The difference may be accentuated by the large
numbers of redshifts in the CIRS catalogs which extend
(non-uniformly) to fainter magnitudes than the simulated catalogs
displayed in D99.  The contrast of the caustics with the background is
unlikely to be a precise cosmological indicator, but it is suggestive
that real clusters more closely resemble the $\Lambda$CDM than the
$\tau$CDM simulated cluster.  

The caustic patterns in the CAIRNS clusters are robust to the addition
of fainter galaxies to the radius-redshift diagrams
\citep{gdk99,cairnsi}.  This result suggests that dwarf galaxies trace
the same caustic pattern as giant galaxies.  This result holds for the
CIRS clusters; in particular, A2199, a CAIRNS cluster, has much deeper
sampling in CIRS (double the redshifts in CAIRNS), yet the mass
profiles agree quite well.  These results imply that any velocity bias
on the scale of infall regions does not depend strongly on galaxy
luminosity.

%The D99 algorithm we use to identify the caustics generally agrees
%with the lines one would draw based on a visual inspection, although the details .  This
%consistency suggests that systematic uncertainties in the caustic
%technique are dominated by projection effects rather than the details
%of the algorithm.  
We discuss some individual clusters in more detail
in $\S \ref{individual}$.

\subsection{Virial and Turnaround Masses and Radii}

The caustic mass profiles allow direct estimates of the virial and
turnaround radius in each cluster.  For the virial radius, we estimate
$r_{200}$ ($r_\Delta$ is the radius within which the enclosed average
mass density is $\Delta\rho _c$, where $\rho _c$ is the critical
density) by computing the enclosed density profile [$\rho (<r) = 3
M(<r)/4\pi r^3$]; $r_{200}$ is the radius which satisfies $\rho
(<r_{200})=200\rho_c$.  In our adopted cosmology, a system should be
virialized inside the slightly larger radius $\sim$$r_{100} \approx
1.3 r_{200}$ \citep{ecf96}.  We use $r_{200}$ because it is more
commonly used in the literature and thus allows easier comparison of
results.  For the turnaround radius $r_{t}$, we use equation (8) of
\citet{rg89} assuming $\Omega _m = 0.3$.  For this value of $\Omega
_m$, the enclosed density is 3.5$\rho _c$ at the turnaround radius.
If the $w$ parameter in the equation of state of the dark energy
($P_\Lambda = w\rho_\Lambda$) satisfies $w\ge -1$, the dark energy has
little effect on the turnaround overdensity
\citep[][]{2002MNRAS.337.1417G}.  Varying $\Omega_m$ in the range
0.02--1 only changes the inferred value of $r_t$ by $\pm$10\%; the
uncertainties in $r_t$ from the uncertainties in the mass profile are
comparable or larger \citep{rines02}.  

Table \ref{radii} lists $r_{200}$, $r_t$, and the masses $M_{200}$ and
$M_t$ enclosed within these radii.  For some clusters, the maximum
extent of the caustics $r_{max}$ is smaller than $r_t$.  For these
clusters, $r_t$ is a lower bound calculated assuming that there is no
additional mass outside $r_{max}$.  The best estimate of the mass
contained in infall regions clearly comes from those clusters for
which $r_{max}\geq r_t$.  The average mass within the turnaround
radius for these clusters is 2.19$\pm$0.18 times the virial mass
$M_{200}$ (the average ratio for all clusters is 1.97$\pm$0.10).
The average turnaround radius is (4.96$\pm$0.08)$r_{200}$
for clusters with $r_{max}\geq r_t$, and (4.75$\pm$0.07)$r_{200}$
averaged over all clusters.  Simulations of the future growth of
large-scale structure
\citep{1991MNRAS.251..128L,2002MNRAS.337.1417G,nl02,busha03} for our
assumed cosmology ($\Omega _m = 0.3, \Omega _\Lambda = 0.7$) suggest
that galaxies currently inside the turnaround radius of a system will
continue to be bound to that system.  In open cosmologies with
$\Omega_\Lambda = 0$, objects in regions where the enclosed density
exceeds the critical density are bound, whereas in closed cosmologies,
all objects are bound to all other objects.  
\citet{busha05} find that the ultimate mass of dark matter haloes
in $\Lambda$CDM simulations is, on average, 1.9 times their masses
$M_{200}$ measured at $z$=0.  Our measure of
$M_t/M_{200}$=2.19$\pm$0.18 confirms their numerical prediction, if we
expect that all the mass within $r_t$ will eventually fall onto the
cluster.
%We would like to compare our results with those of
%\citet{popesso05}, with which we have 30 clusters in common.  However,
%\citet{popesso05} does not publish the mass estimates used in their
%plots so we are unable to make this comparison.  
We compare the virial masses $M_{200}$ from the caustics with masses
calculated using the virial theorem in $\S 7$.

One striking result of this analysis is that the caustic pattern is
often visible beyond the turnaround radius of a cluster.  This result
suggests that clusters may have strong dynamic effects on surrounding
large-scale structure beyond the turnaround radius.  For our assumed
cosmology, this large-scale structure is probably not bound to the
cluster.  See \citet{2005A&A...440L..41P} for an alternative
explanation of this observed structure.

\subsection{Cluster Scaling Relations} %($L_X -\sigma_p$, $M - \sigma _p$)}

Scaling relations between simple cluster observables and masses provide
insight into the nature of cluster assembly and the properties of
various cluster components.  Establishing these relations for local
clusters is critical for future studies of clusters in the distant
universe with the goal of constraining dark energy
\citep{majumdar04,lin04}.  

We apply the prescription of \citet{danese} to determine the mean
redshift $cz$ and projected velocity dispersion $\sigma_p$ of
each cluster from all galaxies within the caustics.  We calculate
$\sigma_p$ using only the cluster members projected within $r_{200}$
estimated from the caustic mass profile.  Note that our estimates of
$r_{200}$ do not depend on $\sigma_p$.

\begin{figure}
\figurenum{8}
\epsscale{0.5}
\plotone{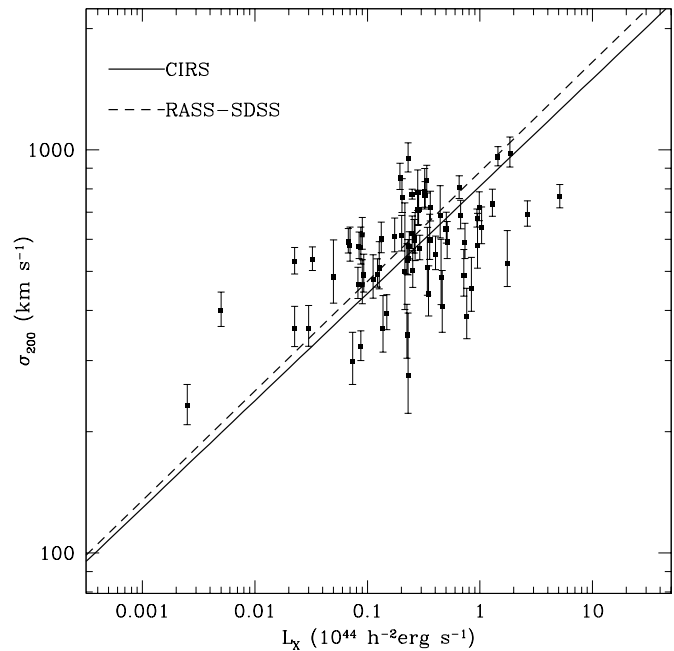} 
\caption{\label{lxsigma} Velocity dispersions at $r_{200}$ compared to X-ray 
luminosities. The solid line is the bisector of the least squares
fits.  The dashed lines show the $\sigma_{200}-L_X$ relations from
RASS-SDSS \citep{popesso05}.  }
\end{figure}

One of the simplest observables of clusters is X-ray luminosity.  The
X-ray luminosities are in the ROSAT band (0.1-2.4 keV) and corrected
for Galactic absorption (taken from REFLEX, NORAS, and BCS/eBCS).
Figure \ref{lxsigma} shows the $L_X -\sigma_p$ relation for the CIRS
clusters along with the best-fit $L_X -\sigma_p$ relation of the
RASS-SDSS \citep{popesso05}.  Although the scatter is large, the
CIRS clusters follow the same relation as the RASS-SDSS sample.
%\subsection{Mass and X-ray Luminosity}

\begin{figure}
\figurenum{9}
\epsscale{0.5}
\plotone{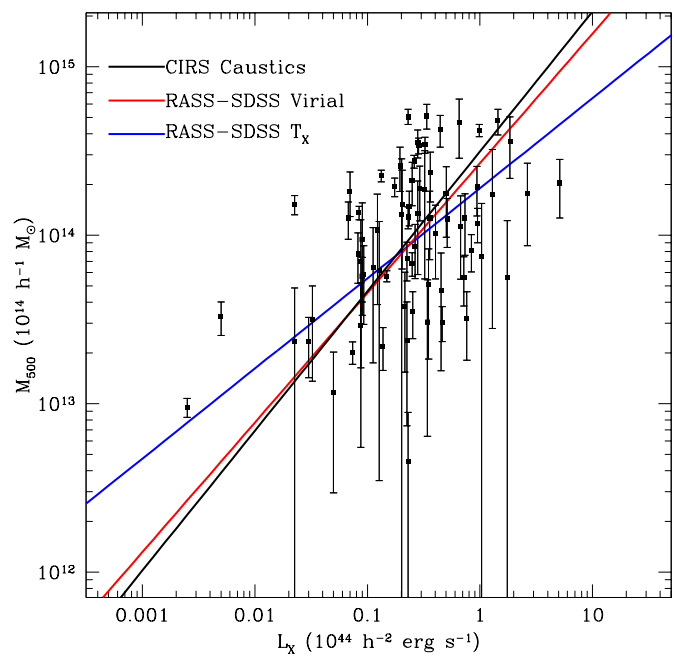} 
%\plotone{f9.eps} 
\caption{\label{clx} Caustic masses at $r_{500}$ compared to X-ray 
luminosities. The solid line is the bisector of the ordinary least squares
fits.  The red and blue lines show the $M_{500}-L_X$ relations for
RASS-SDSS \citep{popesso05} for optical and X-ray masses respectively.  }
\end{figure}

Figure \ref{clx} shows $L_X$ versus $M_{500}$ as estimated from the
caustics.  The bisector of the
least-squares fits to the CIRS sample agrees very well with the
RASS-SDSS $M_{500}-L_X$ relation for masses estimated with the virial
theorem \citep{popesso05}, but both relations are offset from the
$M_{500}-L_X$ relation with masses estimated from $T_X$.  The
significance and origin of this offset merit a more detailed analysis
than possible here. 

The mass-temperature relation
\citep{emn96,1999ApJ...520...78H,2000ApJ...532..694N, frb2001} gives a
straightforward estimate of the mass of a cluster from its X-ray
temperature.  We use the mass-temperature relation rather than, e.g.,
a $\beta$ model to estimate the mass both for simplicity and to ensure
uniformity (X-ray observations allowing more detailed mass estimates
are only available for a small fraction of the CIRS clusters).
Numerical simulations \citep{emn96} suggest that estimating cluster
masses based solely on emission-weighted cluster temperatures yields
similar accuracy and less scatter than estimates which incorporate
density information from the surface brightness profile.

\begin{figure}
\figurenum{10}
\plotone{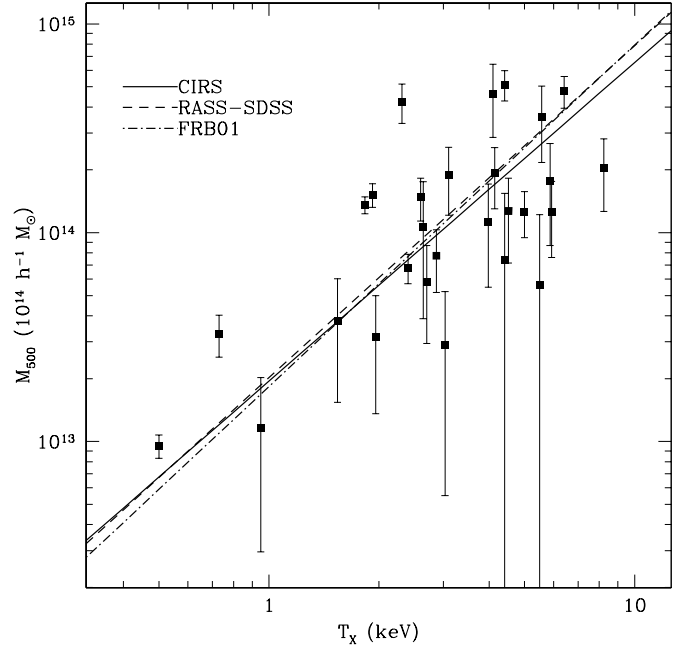} 
\caption{\label{txm5} X-ray temperatures versus caustic masses at $r_{200}$.  The solid line is the bisector of the ordinary least squares
fits.  The dashed and dash-dotted lines are the relations found by
\citet{popesso05} and \citet{frb2001} respectively.  }
\end{figure}

Figure \ref{txm5} shows the X-ray temperature versus $M_{500}$ as
estimates from the caustics for a subset of 28 clusters with X-ray
temperatures listed in \citet{jf99} or \citet{hornerthesis}.  The
solid line is the bisector of the least-squares fits; the other lines
show the best-fit relations of \citet{frb2001} and RASS-SDSS
\citep{popesso05}.  We again find excellent agreement with previously
determined scaling relations.

\begin{figure}
\figurenum{11}
\plotone{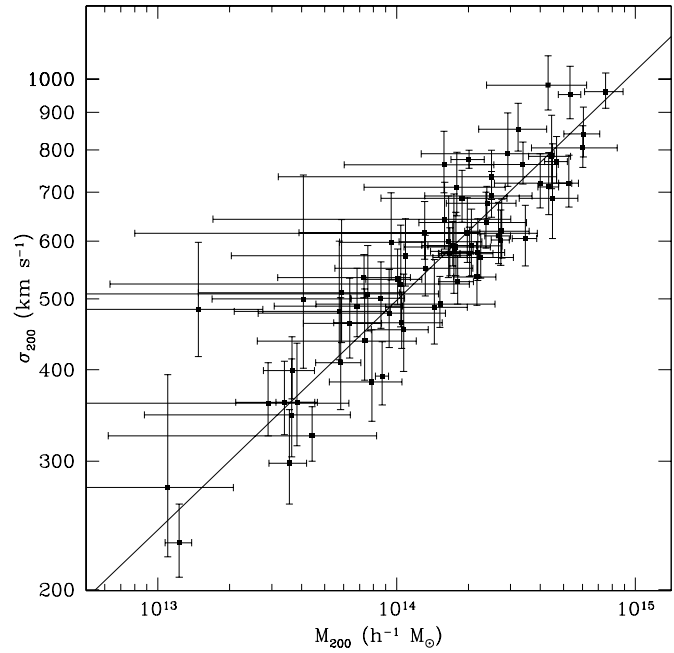} 
\caption{\label{msigma} Caustic masses at $r_{200}$ compared to velocity 
dispersions within $r_{200}$.  The solid line is the bisector of the
ordinary least squares fits.}
\end{figure}

Figure \ref{msigma} shows the $M_{200} - \sigma _p$ relation.  The
tight relation indicates that the caustic masses are well correlated
with velocity dispersion estimates.  The good correlation is perhaps
not surprising because both parameters depend on the galaxy velocity
distribution.  The best-fit slope is
$M_{200}\propto\sigma_p^{3.18\pm0.19}$ with the uncertainty estimated
from jackknife resampling.  We compare the caustic masses to virial
mass estimates in $\S$\ref{virial}.

The excellent agreement between the caustic masses and the X-ray
masses from previously determined scaling relation between mass and
X-ray temperatures confirms the prediction of D99 that the caustic
mass estimate is unbiased.  CAIRNS found similar agreement between
caustic masses and X-ray and virial mass estimates \citep{cairnsi};
\citet{diaferio05} show good agreement between masses estimated from
the caustics and weak lensing.

\subsection{The Shapes of Cluster Mass Profiles \label{shapes}}

We fit the mass profiles of the CAIRNS clusters to three simple
analytic models.  The simplest model of a self-gravitating system is a
singular isothermal sphere (SIS). The mass of the SIS increases
linearly with radius.  \citet{nfw97} and \citet{hernquist1990} propose
two-parameter models based on CDM simulations of haloes.  We note that
the caustic mass profiles mostly sample large radii and are therefore
not very sensitive to the inner slope of the mass profile.  Thus, we
do not consider alternative models which differ only in the inner
slope of the density profile \citep[e.g.,][]{moore99}.  At large
radii, the best constraints on cluster mass profiles come from galaxy
dynamics and weak lensing.  The caustic mass profiles of Coma
\citep{gdk99}, A576 \citep{rines2000}, A2199 \citep{rines02} and the
rest of the CAIRNS clusters \citep{cairnsi} provided strong evidence
against a singular isothermal sphere (SIS) profile and in favor of
steeper mass density profiles predicted by \citet{nfw97} (NFW) and
\citet{hernquist1990}.  Only recently have weak lensing mass estimates
been able to distinguish between SIS and NFW density profiles at large
radii \citep{clowe01,kneib03}.

At large radii, the NFW mass profile increases as ln$(r)$ and the mass
of the Hernquist model converges.  The NFW mass profile is
\beqn
M(<r) = \frac{M(a)}{\mbox{ln}(2) - \frac{1}{2}}[\mbox{ln}(1+\frac{r}{a})-\frac{r}{a+r}]
\eeqn
where $a$ is the scale radius and $M(a)$ is the mass within $a$. We
fit the parameter $M(a)$ rather than the characteristic density
$\delta_c$ (${M(a) = 4\pi \delta_c \rho_c a^3 [\mbox{ln}(2) -
\frac{1}{2}]}$ where $\rho_c$ is the critical density) because $M(a)$
and $a$ are much less correlated than $\delta_c$ and $a$
\citep{mahdavi99}.  The Hernquist mass profile is
\beqn
M(<r) = M \frac{r^2}{(r+a_H)^2}
\eeqn
where $a_H$ is the scale radius and $M$ is the total mass. Note that
$M(a_H) = M/4$. The SIS mass profile is $M(<r)\propto r$.
%\beqn
%M(<r) = M(a=0.5~\Mpc) \frac{r}{0.5~\Mpc}
%\eeqn
We minimize $\chi ^2$ and list the best-fit parameters $a$, $r_{200}$, the
concentration $c_{NFW}$=$r_{200}/a$, and $M_{200}$ for the best-fit NFW
model and indicate the best-fit profile type in Table \ref{mpfitsci}.
We also list the parameter $c_{101}$=$r_{101}/a$; some authors prefer
to use $r_{101}$ as the virial radius.  We perform the fits on all
data points within the maximum radial extent of the caustics $r_{max}$
listed in Table \ref{radii} and with caustic amplitude $\mathcal{A}
\mathnormal{(r)} > 100~\kms$.

Because the individual points in the mass profile are not independent,
the absolute values of $\chi ^2$ are indicative only, but it is clear
that the NFW and Hernquist profiles provide acceptable fits to the
caustic mass profiles; the SIS is excluded for nearly all clusters.
The NFW profile provides a better fit to the data than the Hernquist
profile for 36 of the 72 CIRS clusters; 35 are better fit by a
Hernquist profile and one is best fit by SIS.  A non-singular
isothermal sphere mass profile yields results similar to the SIS;
thus, we report only our results for the SIS.  

\begin{figure}
\figurenum{12}
\plotone{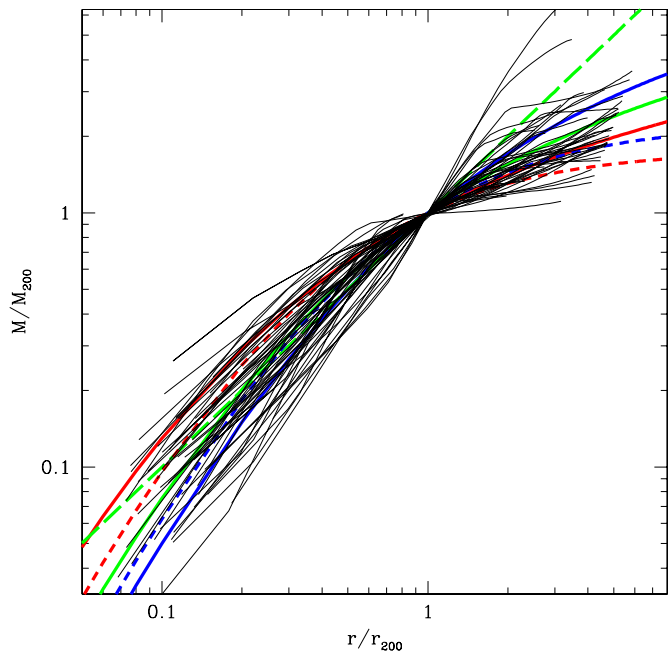} 
\caption{\label{scalem} Scaled caustic mass profiles for the CIRS
clusters compared to simple models. The thin solid lines show the
caustic mass profiles normalized by $r_{200}$ and $M_{200}$.  The
long-dashed line shows a singular isothermal sphere, the colored solid
lines show NFW profiles (with concentrations $c$=3,5,10 from top to
bottom at large radii).  The short-dashed lines are Hernquist profiles
with scale radii different by a factor of two. }
\end{figure}

Figure \ref{scalem} shows the shapes of the caustic mass profiles
scaled by $r_{200}$ and $M_{200}$ along with SIS, NFW, and Hernquist
model profiles.  The colored lines show different model mass profiles.
The straight dashed line is the SIS, the solid lines are NFW profiles
with $c$=3,5, and 10, and the curved dashed lines are Hernquist
profiles with two different scale radii.  The best-fit average profile
is an NFW profile with $c_{200}$=7.2 (this lowers to $c_{200}$=5.2
when the fits are restricted to $r$$\leq$$r_{200}$), consistent with
Table \ref{mpfitsci} and with the values expected from simulations for
massive clusters \citep[NFW, ][]{bullock01}.  All three model profiles
agree fairly well with the caustic mass profiles in the range
(0.1-1)$r_{200}$.  The SIS only fails beyond $\sim$1.5$r_{200}$; this
is why lensing has had trouble distinguishing between SIS and NFW
profiles.  As discussed in D99, the caustic technique can be subject
to large variations for individual clusters due to projection effects.
The best constraints on the shapes of cluster mass profiles are
obtained by averaging over many lines of sight, or for real
observations, over many different clusters.  The current sample is the
largest sample of mass profiles at large radii to date and thus
provides the best possible test of the shapes of cluster mass
profiles.

\begin{figure}
\figurenum{13}
\plotone{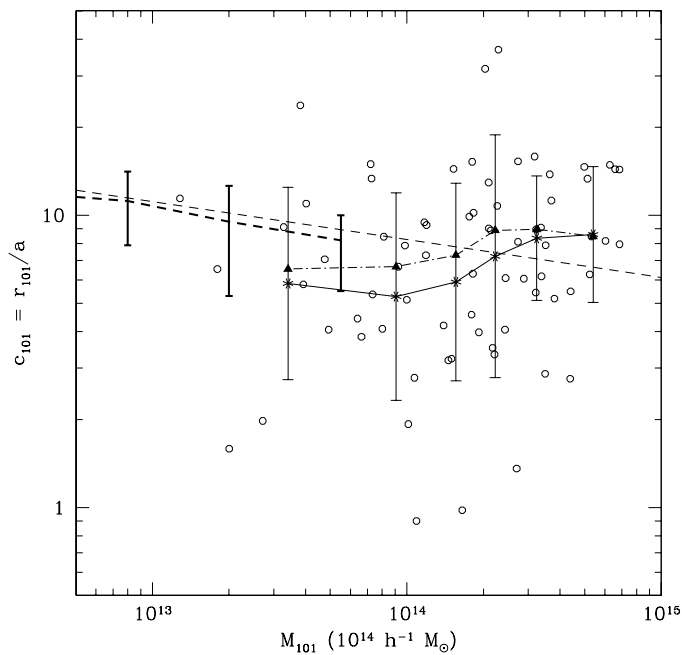} 
\caption{\label{cnfw} Concentrations $c_{101}$=$r_{101}/a$ of the 
best-fit NFW profiles verus $M_{101}$.  The heavy dashed line shows
the expected trend from simulations and the thin dashed line shows a
model of this trend, while the errorbars show the expected $1\sigma$
scatter in $c_{101}$ \citep{bullock01}.  The stars and triangles show the
mean and median values of $\mbox{log}(c_{101})$ in six bins of 12
clusters.  The thin errorbars show the $1\sigma$ scatter in the mean
values of $\mbox{log}(c_{101})$ in each bin.}
\end{figure}

The concentration parameters $c_{200}=r_{200}/a$ for the NFW models
are in the range 2--60, in good agreement with the predictions of
numerical simulations \citep{nfw97,bullock01}.  The differences in $c$
should be small ($\sim$20\%) over our mass range compared to the
scatter in $c$ present in simulated clusters \citep{nfw97,bullock01}.
Figure \ref{cnfw} indicates the average values and $1\sigma$ scatter
of $c_{101}=r_{101}/a$ in simulations \citep{bullock01}.  The dynamic
range of these simulations is not large enough to contain many massive
clusters, but the CIRS clusters agree well with the extrapolation of
the relation found in simulations.  We bin the CIRS clusters into six
bins of 12 clusters and compute the mean and median of
$\mbox{log}c_{101}$.  There is a weak positive correlation of $c$ with
mass (Figure \ref{cnfw}), but the values of $c_{101}$ and the scatter
(thin errorbars) agree well with the model of \citet{bullock01} (the
scatter in CIRS is larger, indicating that observational uncertainties
likely contribute to the observed scatter).  This result addresses one
concern from the CAIRNS mass profiles: the concentrations $c_{200}$
were in the range 5-17 rather than the range 4-6 expected from
numerical simulations for massive clusters
\citep{nfw97,cairnsi}.  Similarly, recent mass profiles from weak
lensing similarly find evidence of high concentrations in A1689
\citep{2005ApJ...619L.143B,2005ApJ...621...53B}, CL0024
\citep{kneib03}, and MS2137 \citep{gavazzi05}.  However, Figure
\ref{cnfw} shows that the CIRS clusters have mass profiles
consistent with those predicted by simulations, although with large
scatter.  If this scatter is physical rather than due to projection
effects in the caustic mass profiles, then the apparent discrepancies
between simulations and observations can be explained by an
unlucky selection of clusters.

\begin{figure}
\figurenum{14}
\plotone{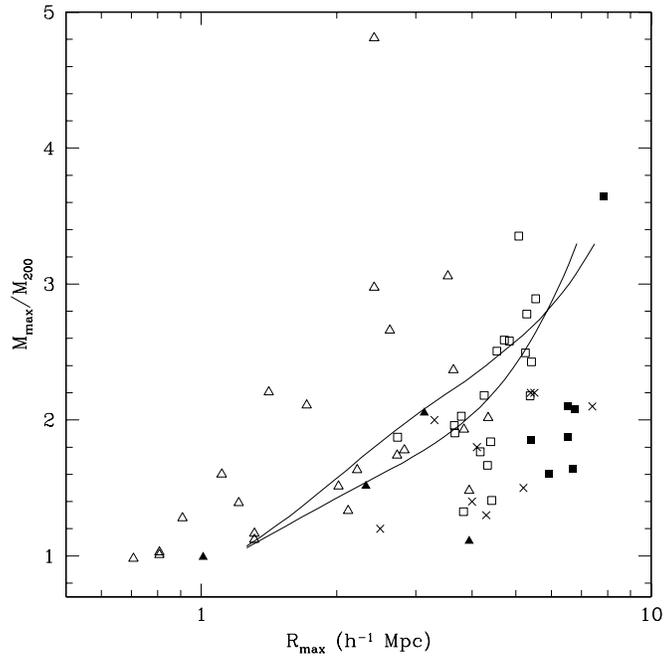}
\caption{\label{ctinker} Ratio of maximum mass $M_{max}$ to virial mass 
$M_{200}$ versus maximum radius $R_{max}$.  Squares are clusters where
$r_t<r_{max}$, triangles are clusters with $r_t\geq r_{max}$.  Solid
points indicate clusters with $M_{200}\geq$3$\times$10$^{14}
h^{-1}M_{\odot}$.  Crosses are the CAIRNS clusters. The two lines
indicate the simulations of \citet{tinker05} for massive clusters for
$\Omega_m=0.1$ and 0.45. }
\end{figure}

Recently, \citet{tinker05} investigated the dependence of the
mass-to-light ratios of large-scale structure on cosmological
parameters.  In particular, they study simulations that reproduce the
observed galaxy angular correlation function and luminosity function.
They then infer the halo occupation distribution (HOD) and conditional
luminosity function (CLF) for several values of $\sigma_8$
($\sigma_8$ is the rms fluctuation in spheres of radius 8$\Mpc$),
yielding a prediction for the dependence of mass-to-light ratio on
halo mass.  Comparing their simulations to observed mass-to-light
ratios of cluster virial regions, \citet{tinker05} concluded that
models with values of $\sigma_8$ and/or $\Omega_m$ smaller than those
found by \citet{tegmark04} provide the best fits.  Using a similar CLF
approach with different parametrization, \citet{2003MNRAS.345..923V}
also conclude that $\sigma_8$ and/or $\Omega_m$ are smaller than
suggested by \citet{tegmark04}.  \citet{tinker05} show that the infall
regions of large-mass halos (those with $M_{200}\geq$3$\times$10$^{14}
h^{-1}M_{\odot}$) in their simulations contain significantly more mass
than the mass profiles inferred from CAIRNS (see their Figure 9).
They suggest that either the caustic masses are systematically
underestimated at large radii or that the observations conflict with
the predictions.  This discrepancy is especially interesting because
it refers to mass ratios and is thus independent of galaxy bias.
Figure \ref{ctinker} shows a similar plot for the CIRS clusters.  The
scatter is large, but the CIRS clusters are in much better agreement
with the predictions of \citet{tinker05} than are the CAIRNS clusters.
However, among clusters with $M_{200}\geq$3$\times$10$^{14}
h^{-1}M_{\odot}$, there does appear to be an offset similar to but
smaller than that of the CAIRNS clusters.  Future comparison of the
caustic mass profiles for the CIRS clusters to clusters in simulations
would help understand this discrepancy (Diaferio et al.~2006, in prep.).

\subsection{The Ensemble CIRS Cluster}

Following other authors \citep[e.g.,][and references
therein]{cye97,bg03,cairnsi}, we construct an ensemble CIRS cluster to
smooth over the asymmetries in the individual clusters.  We scale the
velocities by $\sigma_p$ (Table \ref{sample}) and positions with the
values of $r_{200}$ determined from the caustic mass profiles (Table
\ref{radii}).

\begin{figure}
\figurenum{15}
\plotone{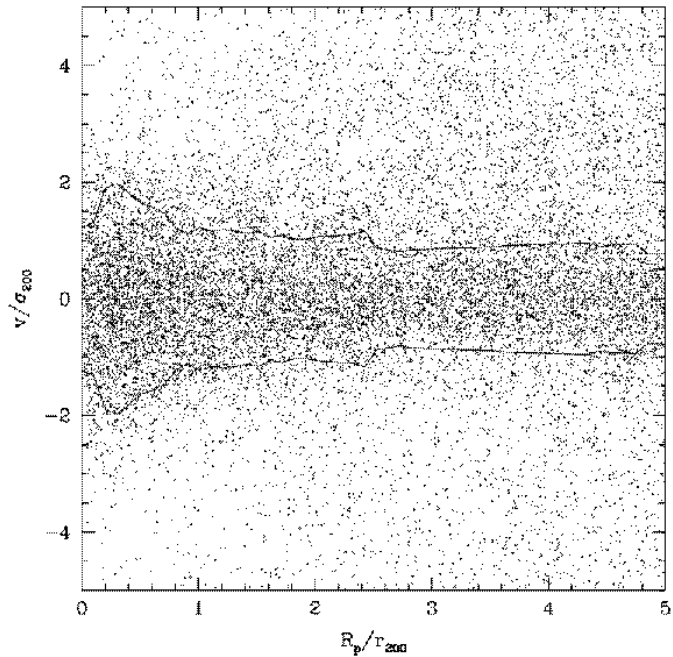} 
\caption{\label{combocaustics} Redshift versus radius for galaxies around
the CAIRNS ensemble cluster. 
%The solid lines indicate our estimate of
%the location of the caustics.  The errorbars are 1-$\sigma$
%uncertainties and are shown only on one side of each caustic for
%clarity. 
}
\end{figure}

Figure \ref{combocaustics} shows the caustic diagrams for the ensemble
cluster.    
%The ensemble
%caustic mass profile yields similar??? results to the individual
%clusters.  At $r_{200}$, the mass is $M_{200} = (xx\pm0.6)
%\sigma_p^2 r_{200}/G$, consistent with the theoretical expectation of
%$M_{200} = 3\sigma_p^2 r_{200}/G$.  The small difference suggests that
%either the measured velocity dispersions are larger than the true
%values (perhaps due to infalling galaxies) or that the estimates of
%$r_{200}$ are low.
The ensemble cluster contains 15103 members within the caustics; 4502
of these are projected within $r_{200}$ and 10,601 have projected radii
between 1 and 5 $r_{200}$ (5$r_{200}$ is comparable to the turnaround
radii in Table \ref{radii}).  These results confirm that infall
regions contain more galaxies than their parent virial regions
\citep{cairnsi,cairnsii}.  The ensemble CIRS cluster contains more
galaxies both within and outside $r_{200}$ than any previous study.

\begin{figure}
\figurenum{16}
\plotone{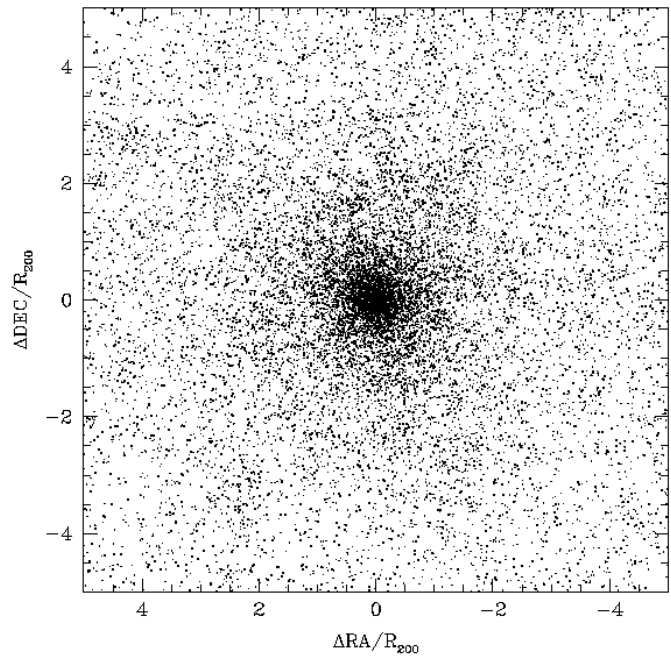} 
\caption{\label{globular} Sky distribution of the ensemble CIRS cluster after removing galaxies near Virgo and NGC4636. }
\end{figure}

Figure \ref{globular} shows the sky distribution of the ensemble
cluster members.  We remove galaxies near Virgo and NGC4636 for this
figure because including them produces non-physical features from the
survey pattern.  Its appearance is reminiscent of globular clusters
and shows that a sufficiently densely sampled ensemble cluster is
close to circularly symmetric.

\section{Comments on Individual Clusters \label{individual}}

Clusters share many common features, but any large sample of clusters
contains some complex systems.  We comment on some of the most
exceptional cases here.

\begin{itemize}

\item {\it A119} This cluster is in the CAIRNS survey.  
See \citet{cairnsi} for discussion.  The mass profile computed here is
based only on SDSS DR4 data which cover only a limited fraction of the
infall region of A119.  Despite this difference, the mass profiles
from CIRS and CAIRNS agree reasonably well.

\item {\it A168} This cluster is in the CAIRNS survey.  
See \citet{cairnsi} for discussion. 

\item {\it NGC 5846/NGC 5813} This group is one of the closest X-ray 
groups.  It has been studied extensively, most recently by
\citet{mahdavi05}, who present a velocity dispersion map.  This map
shows the infall pattern without azimuthal averaging (see their Figure
8).  NGC 5846 is the dominant member of the group, and NGC 5813 is
the center of a merging subgroup.  Despite this ongoing merger, the
velocity field on larger scales shows an infall pattern.  This group
demonstrates that infall patterns can exist in very low-mass systems.

\item {\it A1035} This cluster apparently consists of two systems 
offset by $\sim$3000 $\kms$ in redshift and nearly concentric on the
sky.  There appears to be two infall patterns which overlap at small
radii.  From the hierarchical analysis, the components A and B have
redshifts $z=0.06748$ and $z=0.08008$ respectively.  Component A is
$\sim$10\arcm ~SE of component B.  In the RASS image, A1035 is
elongated NW-SE with a peak in the NW \citep[see Figure 2b of
][]{2003AJ....126.2740L}.  This morphology is confirmed by a 4617
second ROSAT PSPC observation centered 47$\arcm$~SE of A1035.  Our
suggested interpretation is that A1035A and A1035B are both X-ray
clusters separated by $\sim$10\arcm ~and that the more distant A1035B
contributes most of the X-ray flux.  This picture is confirmed by
comparing the RASS image to the SDSS image.  There are two
concentrations of galaxies, one centered roughly on the NW X-ray peak
and one centered in the SE extension.  The brightest several galaxies
in these concentrations confirm that the NW peak is associated with
the higher redshift component and the SE extension with the lower
redshift component.  From the RASS image, A1035B contributes $\sim$2/3
of the total flux, so we estimate $L_{X,A}=7\times10^{42}$erg s$^{-1}$
and $L_{X,B}=2\times10^{43}$erg s$^{-1}$.  If these clusters had been
cleanly resolved by the catalogs, neither would lie above our flux
limit.

\item {\it A1173/A1190} These two clusters are separated by 3.4$\Mpc$ 
in the plane of the sky and by $\lesssim$500 $\kms$ in redshift, they
are likely bound.  A1190 is about twice as luminous as A1173 in
X-rays.  This system is likely a bound binary cluster.  
%RASS shows evidence of more...

\item {\it A1291} This cluster, like A1035A/B, apparently consists 
of two systems offset by $\sim$2000 $\kms$ in redshift and nearly
concentric on the sky (separation $<$4$\arcm$).  The caustic diagram
shows two infall patterns which overlap at small radii.  In addition,
A1318 (not in CIRS) lies $\sim$4$\Mpc$ (in projection) from A1291A/B
at approximately the same redshift as the higher-redshift cluster.
Unlike A1035, A1291 does not contain multiple components in the RASS
image. The peak of the RASS extended source is approximately centered
on a bright galaxy ($M_r\approx$-21.3) in the lower-redshift
component.  It is not clear whether there is any X-ray emission
associated with the higher redshift component.  The total flux is
4.2$\times$10$^{-12}$erg s$^{-1}$cm$^{-2}$, so at most one of the two
components would meet our flux limit if they were resolved.

\item {\it A1728} This cluster has a large offset of 1.3$\Mpc$ between 
the X-ray and hierarchical centers.  The sky distribution of galaxies
reveals two components, one centered roughly on the X-ray peak and the
other 1.3$\Mpc$ WSW which is the hierarchical center (the hierarchical
center is $\sim$10' away from the NED center of ZwCl1320.4+1121, a
cluster without a known redshift).  A caustic plot centered on the
X-ray peak shows an infall pattern with slightly smaller amplitude and
a ``spike'' at the radius of the WSW concentration \citep[similar
spikes at the radii of X-ray groups are seen in A2199;
see][]{rines02}.  The RASS image shows no obvious X-ray emission from
the WSW concentration.  If we apply the caustic technique to the
pattern centered on the X-ray peak, the total mass is smaller by about
a factor of two.

\item{\it A1750A/B} As noted in $\S$\ref{xcs},  the X-ray emission of A1750 
has a secondary component to the south of the central emission.  These
components are separated by 340$\kpc$ in projection and
$\sim$1300 $\kms$ in radial velocity \citep{belsole04}.  The
hierarchical center is located close to the primary X-ray source in
both position and redshift.

\item {\it A2061/A2067} These two clusters are separated by 1.8$\Mpc$ 
in the plane of the sky and by $\sim$600 $\kms$ in redshift, they are
likely bound.  A2061 is about four times more luminous in X-rays than
A2067, so this system is similar to the A2199 supercluster, i.e., a
central, massive cluster with an infalling group.  These clusters are
part of the Cor Bor supercluster \citep{small98,marini04}.  A2061
contains an X-ray 'plume' extending in the direction of A2067,
suggesting a dynamical connection between the two systems
\citep{marini04}.  The A2069 supercluster (not in CIRS) lies behind
both A2067 and A2061 \citep{1988AJ.....95..267P,small98}. It is
possible that hot gas in condensed substructures in the A2069
supercluster contribute to the measured X-ray flux attributed to A2067
and/or A2061.  

\item {\it A2149} This cluster has two components along the line of 
sight, one at $z$=0.0675 and one at $z$=0.1068.  The former redshift
is adopted by eBCS and the latter redshift is adopted by NORAS.  A
close inspection of the RASS image for this cluster shows an extended
area of X-ray emission peaked on the BCG of the $z$=0.0675 component.
We conclude that most of the X-ray flux comes from the system at lower
redshift with possible contamination from the more distant system.

%\item {\it A2192/NGC 6159} A2192 has an incorrect redshift listed in 
%NORAS.  The correct redshift is $z=0.18$ \citep{}.  The redshift
%listed in NORAS is that of the NGC 6159 group dicussed in
%\citet{rines02}.  This group is on the outskirts of A2199.
%Surprisingly, it shows an infall pattern of its own which extends
%until the caustics of A2199 overwhelm it.

\item {\it A2197/A2199} This cluster is in the CAIRNS survey.  
See \citet{rines01b,rines02} for discussion.  A2197 has a 750$\kpc$
offset between the X-ray and hierarchical centers.  A2197 is composed
of two X-ray groups, A2197W and A2197E.  The X-ray center is the peak
of A2197W; the hierarchical center is located approximately midway
between A2197W and A2197E.

\item {\it A2244/A2245/A2249} These three clusters are in a fairly 
small volume of the universe.  A2245 ($z$=0.086) is the most massive
of the three, followed by A2244 ($z$=0.0997) $\sim$30\arcm north.
A2249 ($z$=0.086) is centered about 2 degrees east and just off the
edge of the SDSS spectroscopic survey.  In addition, there's a loose
system (A2241B, in the background of A2241) SW of A2245 which is at
$z$$\sim$0.097 but seems to have multiple components along the line of
sight.

\end{itemize}

\section{Comparison to Virial and Projected  Mass Estimates}

\citet{zwicky1933,zwicky1937} first used the virial theorem to estimate the
mass of the Coma cluster.  With some modifications, notably a
correction term for the surface pressure \citep{1986AJ.....92.1248T},
the virial theorem remains in wide use \citep[e.g.,][and references
therein]{girardi98}.  Jeans analysis incorporates the radial
dependence of the projected velocity dispersion \citep[e.g.,][and
references therein]{cye97,2000AJ....119.2038V,bg03} and obviates the
need for a surface term.

Jeans analysis and the caustic method are closely related.  Both use
the phase space distribution of galaxies to estimate the cluster mass
profile.  The primary difference is that the Jeans method assumes that
the cluster is in dynamical equilibrium; the caustic method does not.
The Jeans method depends on the width of the velocity distribution of
cluster members at a given radius, whereas the caustic method
calculates the edges of the velocity distribution at a given radius.
The caustic method is not independent of the Jeans method, as the D99
method minimizes $|<$$v_{esc}^2$$>_R - 4 <$$v^2$$>_R|$ within the
virial region with radius R (see D99 for a more detailed discussion).
Mass estimates based on Jeans analysis thus provide a consistency
check but not an independent verification of the caustic mass
estimates.

Applying the Jeans method requires an assumption about either the mass
distribution or the orbital distribution.  Typically, one assumes that
light traces mass and thus that the projected galaxy density is
proportional to the projected mass density \citep[e.g.,][]{girardi98}
or one assumes a functional form for the orbital distribution
\citep[e.g.,][]{bg03}.  Note that most authors make the implicit
assumption that the orbital distribution of the dark matter can be
inferred from that of the galaxies.  Many recent simulations suggest
that little ($\sim$10\%) or no velocity bias exists on linear and
mildly non-linear scales
\citep{kauffmann1999a,kauffmann1999b,diemand04,faltenbacher05}.
However, galaxies in simulated clusters often have significantly
different orbital distributions than the dark matter
\citep[e.g.,][]{1999ApJ...516..530K,dkthesis,faltenbacher05}. Dynamical
friction should produce a smaller velocity dispersion for cluster
galaxies than dark matter
\citep{1999ApJ...516..530K,2003ApJ...590..654Y,2005MNRAS.361.1203V},
although they may undergo more frequent mergers (due to their smaller
velocities) or tidal disruption, resulting in a larger velocity
dispersion of the surviving cluster galaxies relative to the dark
matter \citep{1999ApJ...523...32C,diemand04,faltenbacher05}.  Note
also that velocity bias may depend on the mass of the cluster
\citep{berlind03} or the luminosities of the tracer galaxies
\citep{2006MNRAS.366.1455S}.  An important caveat is that the subhalo
distribution of cluster galaxies in simulations does not match the
observed distributions of cluster galaxies: real cluster galaxies
trace the dark matter profile, while simulated cluster galaxies are
significantly antibiased in cluster centers, likely due to the
resilience of the stellar component of a subhalo against tidal
disruption relative to the total subhalo \citep[e.g.,][]{diemand04,
faltenbacher06}.  The details of velocity bias depend also on the
selection used to identify subhaloes, e.g., whether subhalo mass is
measured prior to or after a subhalo enters the cluster
\citep{faltenbacher06}.  The amount of velocity bias in simulations is
typically $\sim$10\% for cluster-size haloes, although disagreement
remains on whether this bias is positive or negative.  Velocity bias
of this size would lead to virial masses overestimated or
underestimated by $\sim$20\%.  Future work is clearly needed to
understand the nature and significance of velocity bias for cluster
mass estimates.  There is no clear evidence from observations in favor
of or against velocity bias in clusters, although the generally good
agreement between X-ray, lensing, and virial mass estimates suggests
that any velocity bias is not large \citep[$\lesssim$20\%;
e.g.,][]{girardi98,popesso04,diaferio05}.  Because significant
velocity bias would produce incorrect virial mass estimates, the
agreement between virial and caustic mass estimates is no guarantee
that the caustic mass estimates are accurate.  However, the mechanisms
which cause velocity bias are less effective in the outskirts of
clusters, so the caustic technique should be less affected by velocity
bias than estimates based on Jeans' analysis.  Analysis of simulations
of large-scale structure with very large dynamic range
\citep{springel05} or of individual clusters \citep{borgani06} may
provide a clearer understanding of the potential impact of velocity
bias on cluster mass estimates from galaxy kinematics.

Another common assumption is that the kinematics of galaxies are
independent of their luminosities.  The consistency of CAIRNS mass
profiles and velocity dispersion profiles for $K_s$-band
luminosity-limited and deeper samples indicates that this is a
reasonable approximation \citep{cairnsii}, but theoretical models of
subhaloes suggest that velocity bias should depend on luminosity
\citep{2006MNRAS.366.1455S}.  We plan to use the CIRS sample to test
this assumption in future work by studying the luminosity dependence
of the kinematic distribution of galaxies.  Note that the CIRS
clusters are sampled to depths of $\sim$$M^*$+1 for the most distant
clusters to $\sim$$M^*$+7 for Virgo, so the CIRS caustics sample a
wide range of luminosity limits.  We confirm that the ratio of virial
to caustic mass estimates in an X-ray luminosity-limited sample shows
no significant trend with redshift, indicating that the different
spectroscopic sampling limits do not affect these comparisons.
Similarly, the ratio of caustic masses to X-ray luminosities is not
significantly different for an X-ray luminosity-limited sample.  These
results suggest that any velocity bias does not depend strongly on
galaxy luminosity.  We defer further discussion to future work.

We apply the virial mass and projected mass estimators
\citep{htb} to the CIRS clusters.  For the latter, we assume the
galaxies are on isotropic orbits.  We must define a radius of
virialization within which the galaxies are relaxed.  We use $r_{200}$
(Table \ref{radii}) and include only galaxies within the caustics.  We
thus assume that the caustics provide a good division between cluster
galaxies and interlopers (see Figures \ref{allcirs1}-\ref{allcirs6}).

We calculate the virial mass according to
\begin{equation}
M_{vir} = \frac{3 \pi}{2} \frac{\sigma_p^2 R_{PV}}{G}
\end{equation}
where $R_{PV} =  2N(N-1)/\sum_{i,j>i}R_{ij}^{-1}$ is the projected
virial radius and $\sigma_p^2 = \sum_i (v_i-\bar v)^2/(N-1)$. 
If the system does not lie entirely within $r_{200}$, a surface
pressure term 3PV should be added to the usual virial theorem so that
$2T + U = 3PV$. The virial mass is then an overestimate of the mass
within $r_{200}$ by the fractional amount 
\begin{equation}
\label{virialc}
C =  4\pi r_{200}^3 \frac{\rho (r_{200})}{\int_{0}^{r_{200}}4\pi r^2 \rho dr} \Bigg[{\frac{\sigma _{r} (r_{200})}{\sigma (<r_{200})}\Bigg]^2}
\end{equation}
where $\sigma_r (r_{200})$ is the radial velocity dispersion at
$r_{200}$ and $\sigma (<r_{200})$ is the enclosed total velocity
dispersion within $r_{200}$ \citep[e.g.,][]{girardi98}.  In the
limiting cases of circular, isotropic, and radial orbits, the maximum
value of the term involving the velocity dispersion is 0, 1/3, and 1
respectively. 

The projected mass estimator is more robust in the presence of close
pairs.  The projected mass is
\begin{equation}
M_{proj} = \frac{32}{\pi G} \sum_i{R_i (v_i-v)^2}/N
\end{equation}
where we assume isotropic orbits and a continuous mass distribution.
If the orbits are purely radial or purely circular, the factor 32
becomes 64 or 16 respectively.  We estimate the uncertainties using
%the jackknife technique and 
the limiting fractional uncertainties $\pi^{-1} (2 \mbox{ln}
N)^{1/2}N^{-1/2}$ for the virial theorem and $\approx 1.4 N^{-1/2}$
for the projected mass.  These uncertainties do not include systematic
uncertainties due to membership determination or the assumption of
isotropic orbits in the projected mass estimator.
%\citet{dkthesis} applies the
%virial theorem to simulated clusters using the velocities of fewer
%than 100 ``galaxies'' per cluster (other authors test the accuracy of
%the virial mass estimator assuming the existence of unphysically
%large numbers of test particles) and finds that mass estimates for
%individual clusters have a scatter of $\sim$35\%, comparable to the
%scatter in the caustic estimate found in $\S 7.1$.
%We adopt the larger of the two uncertainty estimates. 
Table \ref{virial} lists the virial and projected mass estimates. 

\begin{figure}
\figurenum{17}
%\plottwo{cirs.mcomp.eps}{cirs.mcomp2.eps} 
\plotone{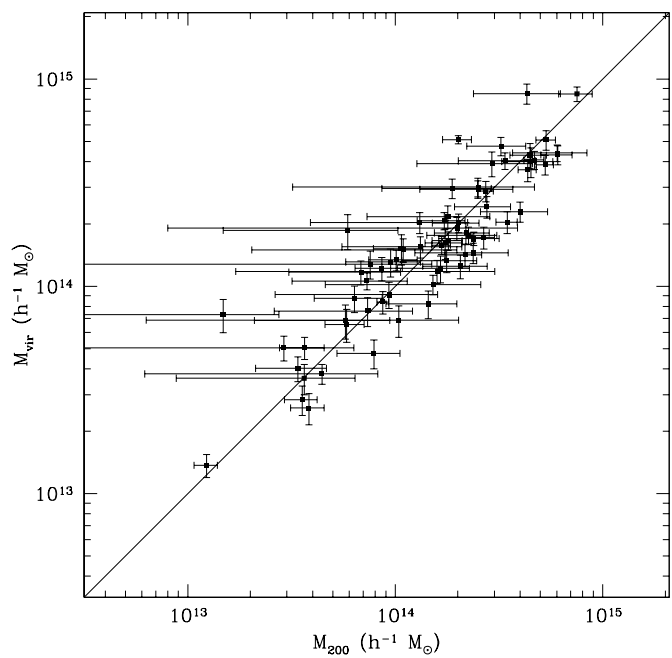} 
\caption{\label{cirsvc} Caustic masses at $r_{200}$ compared to 
virial masses at the same radius.  Errorbars show 1$\sigma$
uncertainties and the solid line has slope unity. 
%(a) Linear scale (b) Logarithmic scale. 
}
\end{figure}

\begin{figure}
\figurenum{18}
\epsscale{1.0}
\plotone{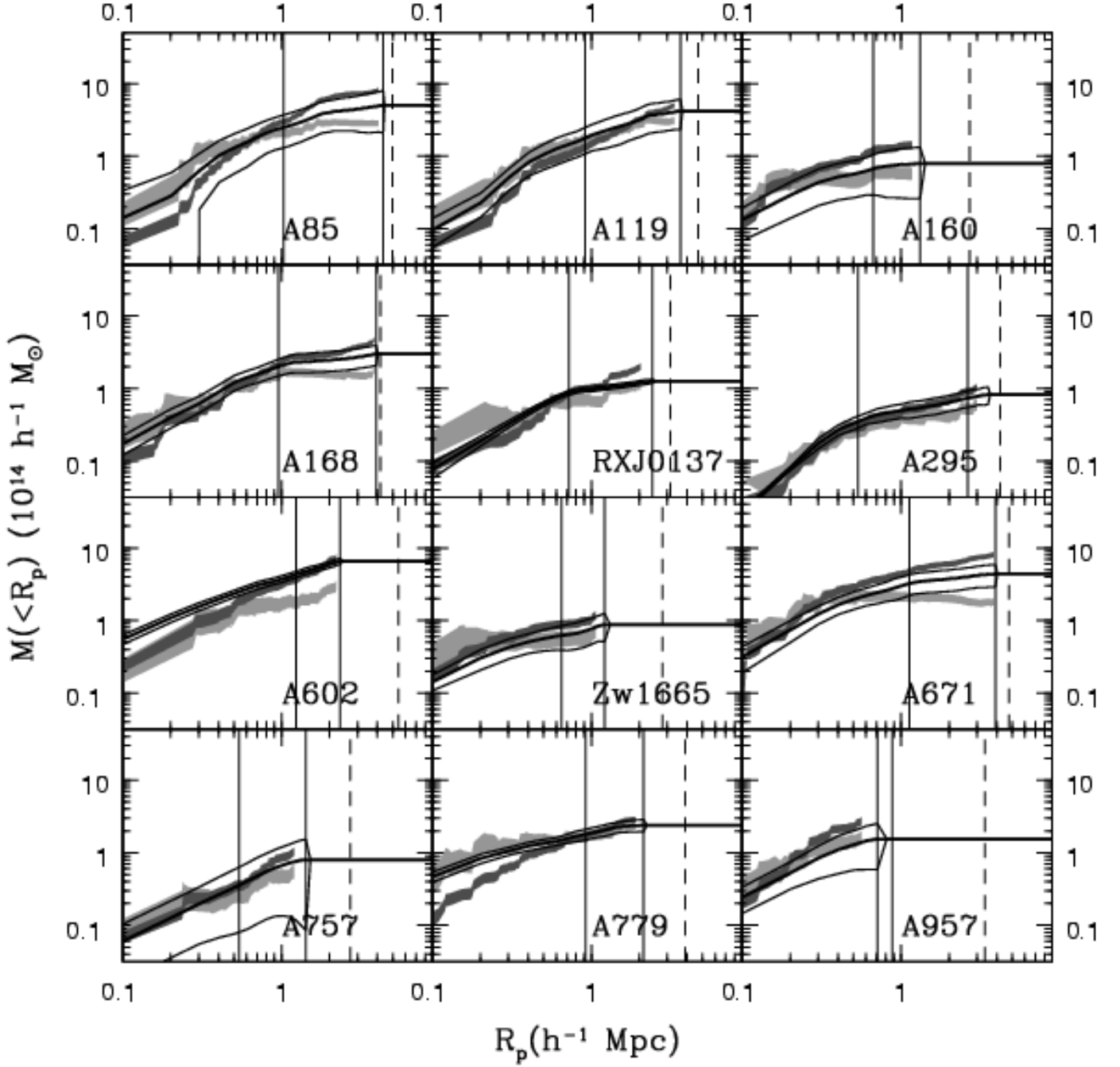}
\caption{\label{allcirsm1} Comparison of caustic mass profiles to those 
estimates from the virial theorem and the projected mass estimator.
The thick solid lines show the caustic mass profiles and the thin
lines show the 1$\sigma$ uncertainties in the mass profiles. The axes
are identical in all panels. The vertical bars indicate $r_{200}$ and
the maximum radius of the caustic mass profile (the smaller of
$r_{max}$, the extent of the infall pattern, and $r_t$, the turnaround
radius).  Vertical dashed lines indicate $r_t$ for clusters where the
infall pattern truncates before $r_t$. Red and green shaded regions
show the formal 1$\sigma$ uncertainties in the virial and projected
mass profiles. Figures \ref{allcirsm2}-\ref{allcirsm6} show similar
plots for the rest of the sample.}
\end{figure}

\begin{figure}
\figurenum{19}
\plotone{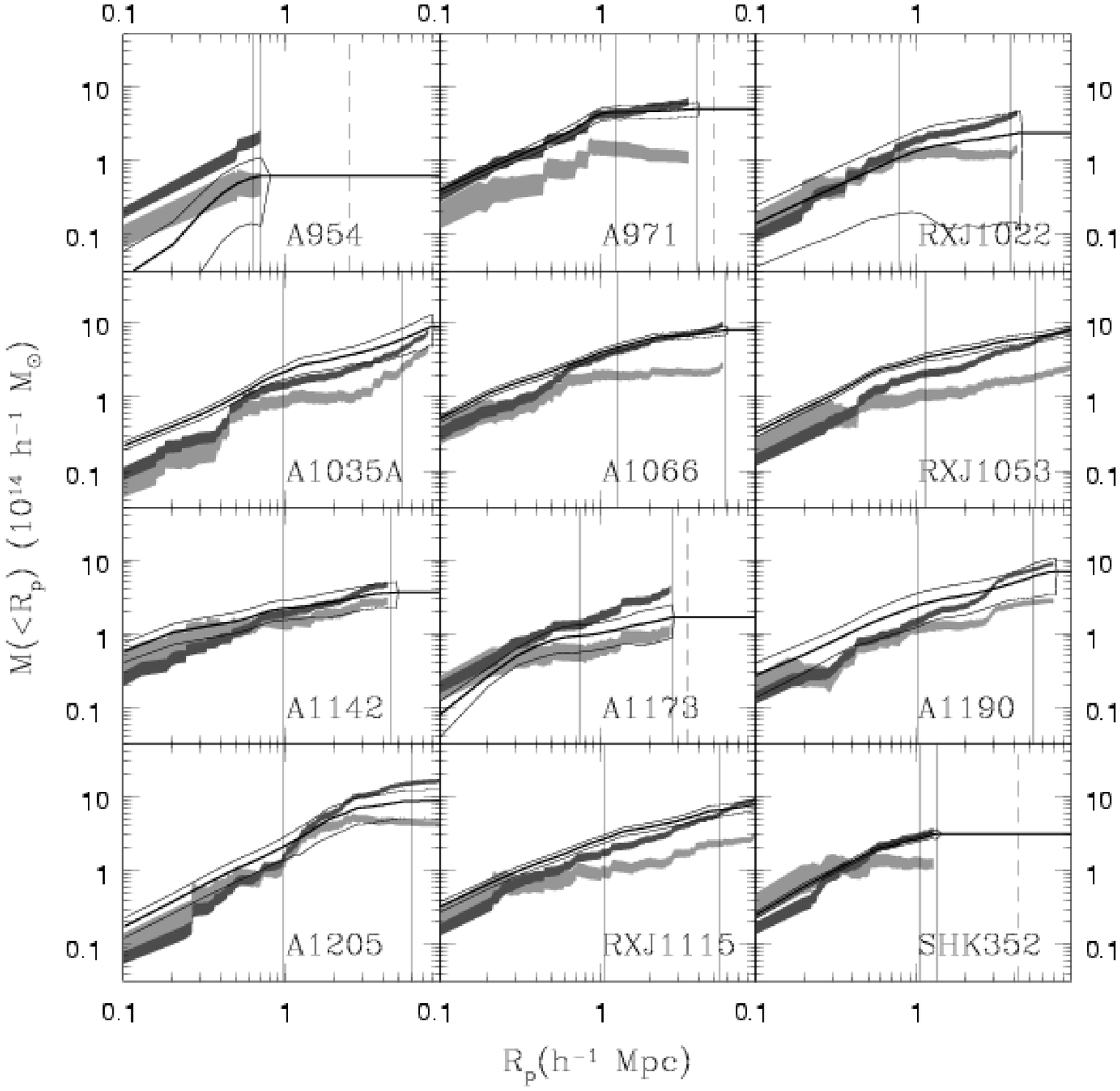}
\caption{\label{allcirsm2} See Figure \ref{allcirsm1}.}  
\end{figure}

\begin{figure}
\figurenum{20}
\plotone{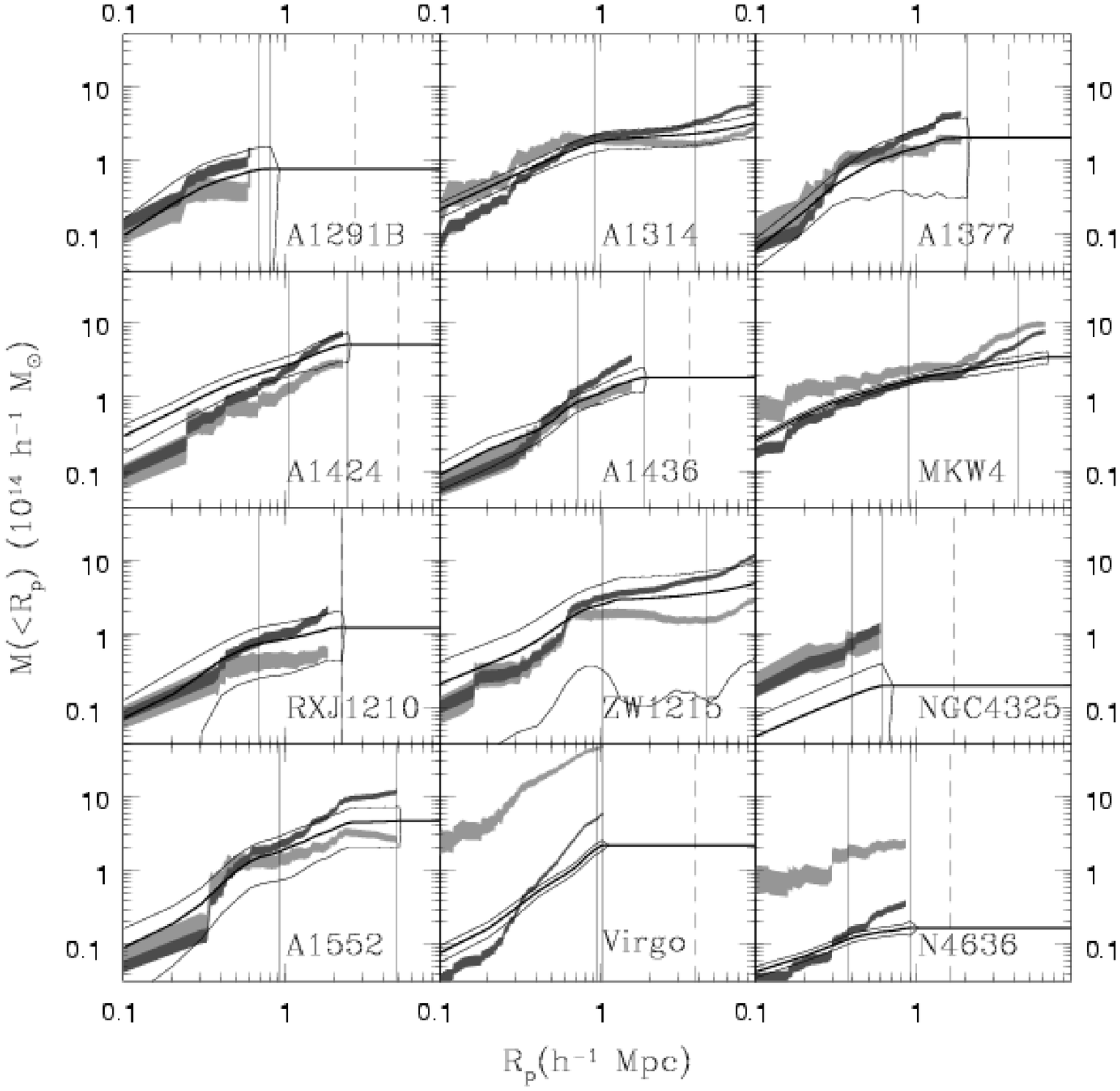}
\caption{\label{allcirsm3} See Figure \ref{allcirsm1}.}  
\end{figure}

\begin{figure}
\figurenum{21}
\plotone{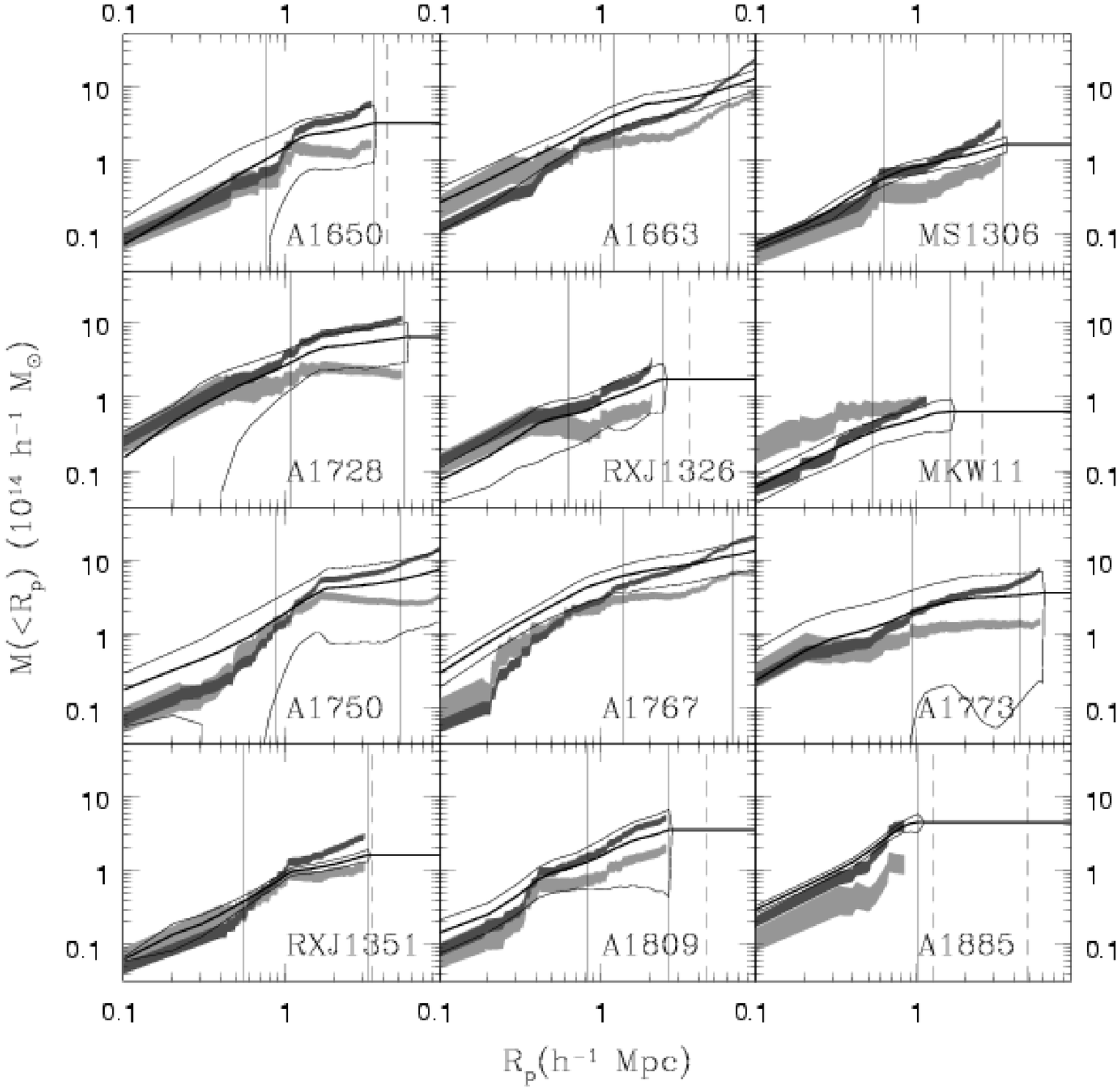}
\caption{\label{allcirsm4} See Figure \ref{allcirsm1}.}  
\end{figure}

\begin{figure}
\figurenum{22}
\plotone{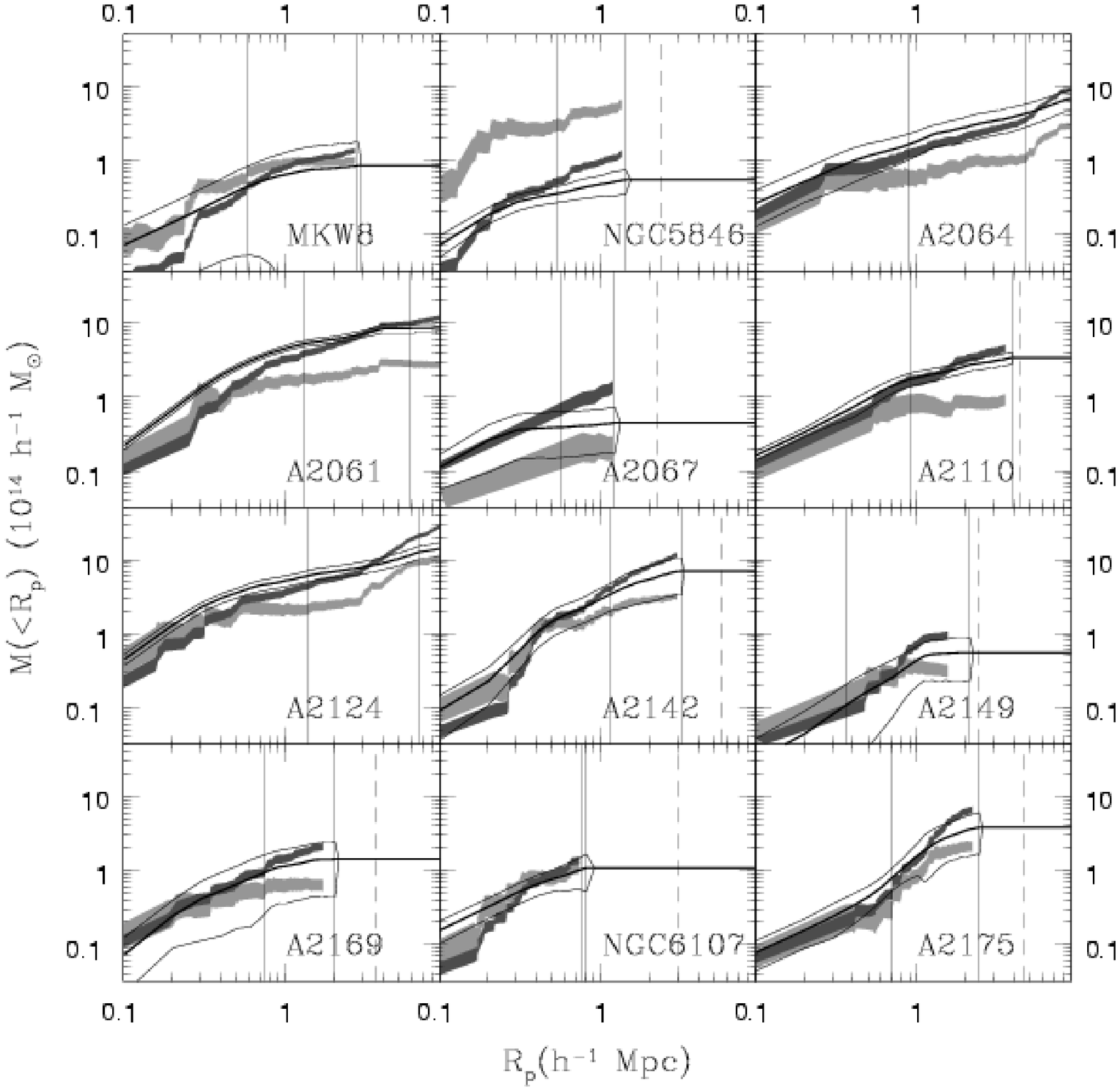}
\caption{\label{allcirsm5} See Figure \ref{allcirsm1}.}  
\end{figure}

\begin{figure}
\figurenum{23}
\plotone{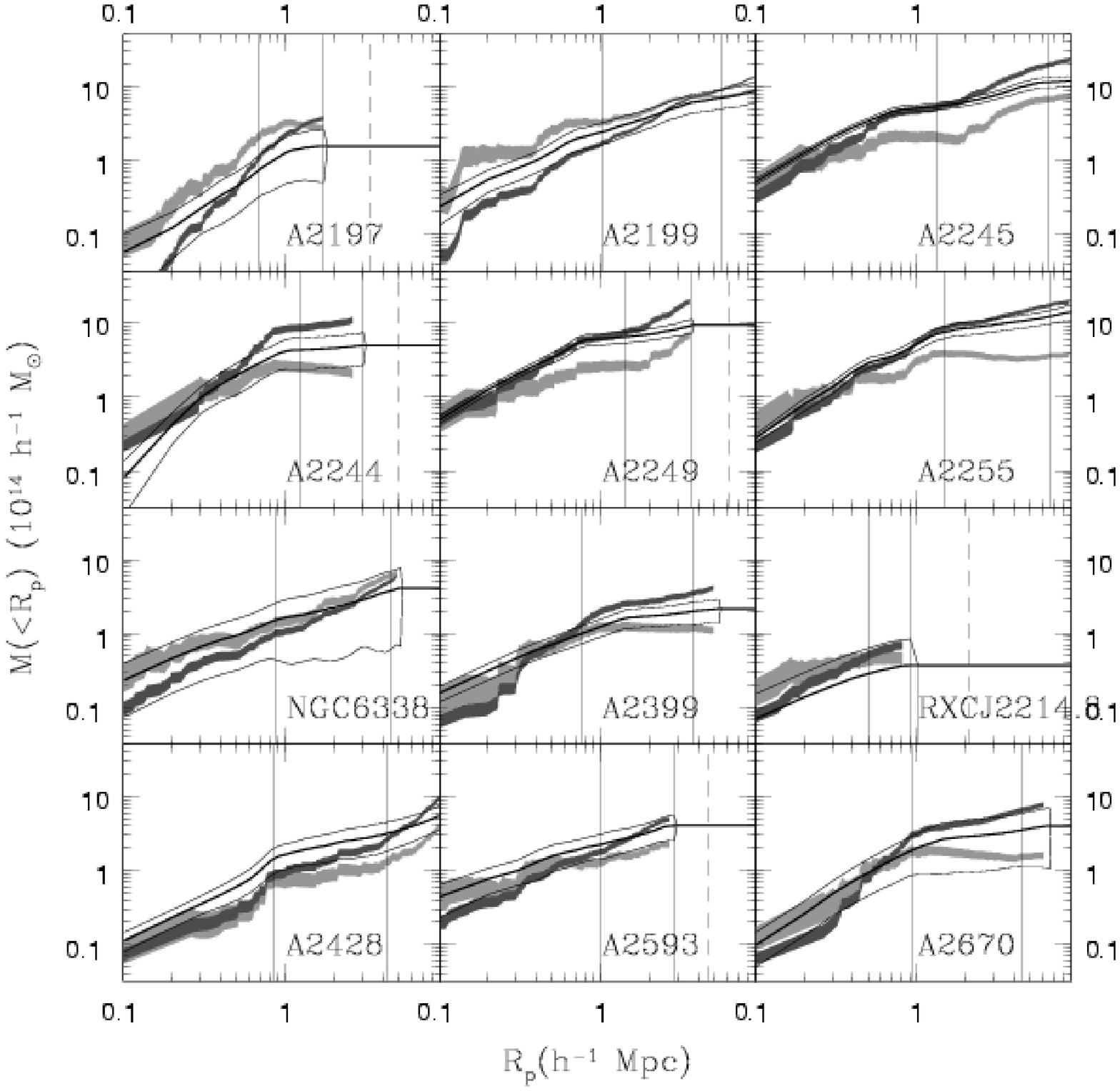}
\caption{\label{allcirsm6} See Figure \ref{allcirsm1}.}  
\end{figure}

%Figure \ref{vp} compares the projected mass estimates to the virial
%mass estimates.  The two mass estimators generally give consistent
%results, but the uncertainties are probably underestimated because
%they do not include systematic uncertainties.  The error-weighted mean
%ratio of the estimates is $M_p/M_v = 1.18 \pm 0.05$.  This result is
%not dominated by the outlier, A539.  If we eliminate this data point,
%the mean ratio is $1.16\pm 0.05$.  Similarly, 
Figure \ref{cirsvc}  compare the virial and caustic mass estimates at
$r_{200}$.  The mean ratios of these estimates are
$M_c/M_v = 1.01 \pm 0.04$. %and $M_c/M_p = 0.xx \pm 0.05$.  
The caustic
mass estimates are consistent with virial mass estimates even assuming a
correction factor $C\approx 0.1-0.2 M_{vir}$, consistent with the best-fit
NFW profiles \citep[see also ][]{cye97,girardi98,kg2000,cairnsi}.

Figures \ref{allcirsm1}-\ref{allcirsm6} compare the mass profiles
estimated from the caustics, virial theorem, and projected mass
estimator.  In the caustic technique, errors on the mass profiles are
estimated by the inverse of the peak of the galaxy number density in
the redshift diagrams. The errors derived with this recipe agree with
the typical spread of mass profiles measured in simulated clusters
(D99, Diaferio et al 2006, in prep).  The projected mass estimator
consistently overestimates the mass at small radii and underestimates
the mass at large radii relative to the other profiles.  This behavior
suggests that this estimator is best for estimating virial masses but
not mass profiles.  The virial and caustic mass profiles generally
agree although there are many clusters with large disagreements.  The
caustic mass profiles do not appear to consistently overestimate or
underestimate the mass relative to the virial mass profiles.  This
result supports our use of caustic mass profiles as a tracer of the
total cluster mass profile ($\S$\ref{shapes}).

\section{Velocity Dispersion Profiles}

Several authors have explored the use of the velocity dispersion
profile (VDP) of clusters as a diagnostic of their dynamical states.
For example, \citet{1996ApJ...473..670F} find that VDPs typically have
three shapes: increasing, flat, or decreasing with radius.

\begin{figure}
\figurenum{24}
\plotone{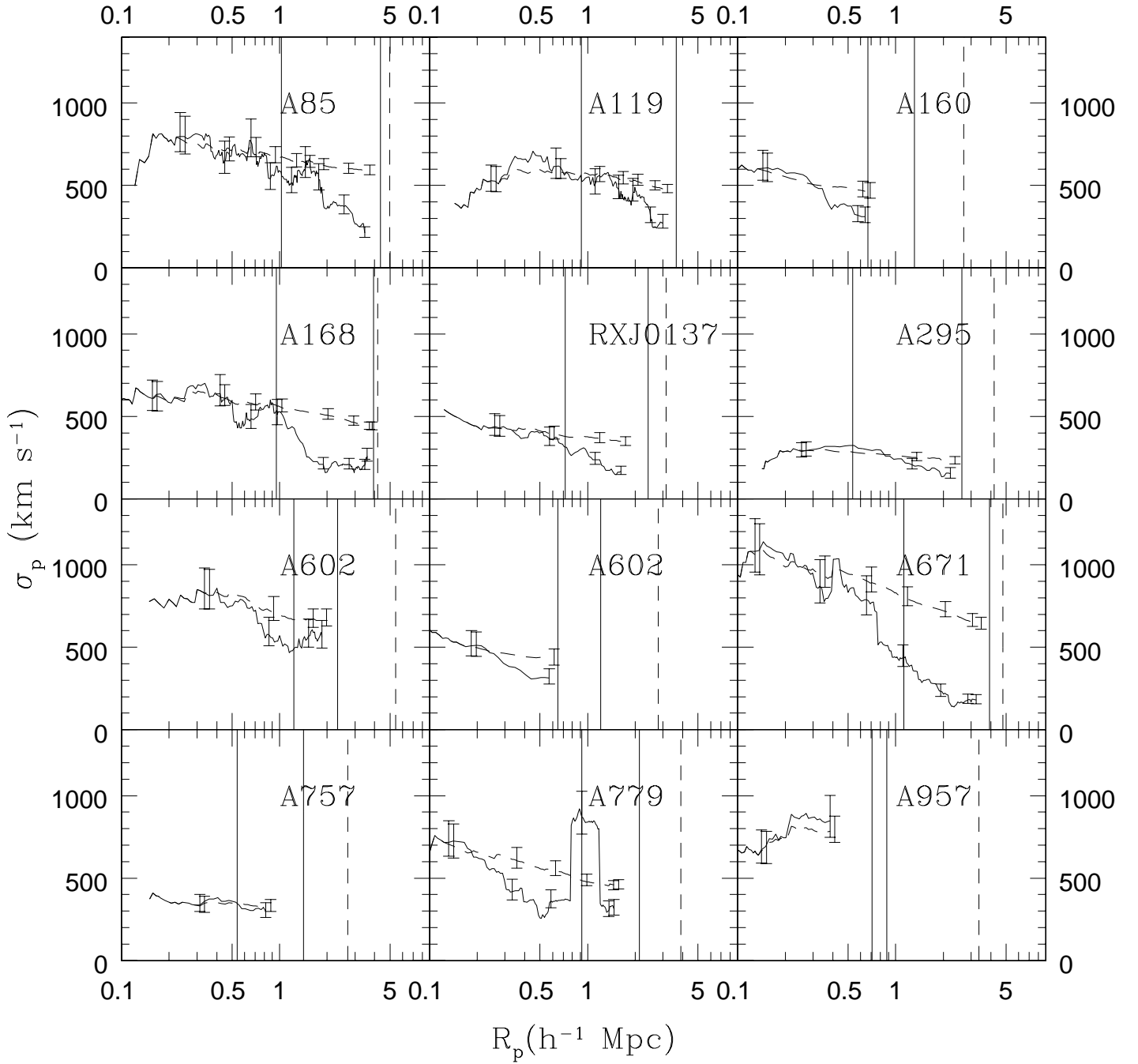}
\caption{\label{allcirss1} Velocity dispersion profiles for the CIRS
clusters. The  solid lines show the velocity
dispersion profile of member galaxies (those within the caustics) in a moving bin of 25 galaxies.  Errorbars indicate 1-$\sigma$ uncertainties for independent bins.  The  dashed lines show the enclosed velocity dispersion profiles.  
%The dash-dotted lines show the VDPs
%predicted by the Hernquist mass models which best fit the caustic mass
%profiles (assuming isotropic orbits).  
Vertical lines 
indicate $r_{200}$ and $r_t$. The axes are identical in all panels. 
%Figures
%\ref{allcirss2}-\ref{allcirss6} show similar plots for the rest of the
%sample.
}
\end{figure}

We calculate the VDPs of the CIRS clusters using all galaxies inside
the caustics.  Most of the CIRS clusters display decreasing VDPs
within about $r_{200}$ (Figure \ref{allcirss1} displays the first 12
clusters in RA).  The VDPs either flatten out or continue to decrease
between $r_{200}$ and $r_t$, consistent with the results of CAIRNS.

%%Figure \ref{combovdp}
%%shows the  VDP of the ensemble cluster. 
%in good agreement.  This agreement suggests that the caustic mass
%profiles are consistent with Jeans analysis (see $\S 7.2$) and that
%the orbits are not far from isotropic, even outside the virial radius.
%However, this result could perhaps be mimicked by a population of
%infalling galaxies not in equilibrium and not on radial orbits.
%Indeed, many authors find evidence of an infalling population
%dominated by blue or emission line galaxies with larger velocity
%dispersions than the red and presumably more relaxed galaxies
%\citep{mohr96,1997ApJ...476L...7C,mahdavi99,kg2000,ellingson01}.

Some authors \citep{cye97,girardi98} suggest that an accurate estimate
of $r_{200}$ can be obtained for a cluster from the asymptotic value
of the enclosed velocity dispersion $\sigma _p (<r)$ calculated for
all galaxies within a given radius.  Many of the CIRS clusters,
however, display no obvious convergence in the enclosed velocity
dispersion (shown by dashed lines in Figure \ref{allcirss1}).  If the
rich clusters in the CNOC1 survey are similar to their CIRS cousins,
the use of the asymptotic value of the enclosed velocity dispersion to
estimate $r_{200}$ may be unreliable.  The caustic technique provides
an alternative method for estimating $r_{200}$, although applying it
requires many more redshifts than are needed for computing the
velocity dispersion.

\section{Summary}

We use the Fourth Data Release of the Sloan Digital Sky Survey to test
the ubiquity of infall patterns around galaxy clusters.  The CAIRNS
project found infall patterns in all 9 clusters in the survey, but the
cluster sample was small and incomplete.  Matching X-ray cluster
catalogs with SDSS, we search for infall patterns in a complete sample
of \ncirs X-ray selected clusters.  Well-defined infall patterns are
apparent in most of the clusters, with the fraction decreasing with
increasing redshift due to shallower sampling.  All clusters in a
well-defined sample limited by redshift (ensuring good sampling) and
X-ray flux (excluding superpositions) show infall patterns sufficient
for calculating caustic mass profiles.
%This high success rate is better than predicted by simulations.  
Similar to CAIRNS, cluster infall patterns are better defined in observations
than in simulations.  Further work is needed to determine whether this
difference lies in the galaxy formation recipes used in simulations or
is more fundamental.  

We use the infall patterns to compute mass profiles for the clusters
and compare them to model profiles.  We confirm the conclusion of
CAIRNS that cluster infall regions are well fit by NFW and Hernquist
profiles and poorly fit by singular isothermal spheres.  Observed
clusters resemble those in simulations, and their mass profiles are
well described by extrapolations of NFW or Hernquist models out to the
turnaround radius.  The scaled mass profiles indicate that NFW
profiles are favored over Hernquist and SIS profiles.  The shapes of
the best-fit NFW cluster mass profiles agree reasonably well with the
predictions of simulations; the average mass profile has
$c_{200}$$\approx$7.2, slightly larger than the value for cluster size
halos extrapolated from \citet{bullock01} and with similar scatter.
These mass profiles test the shapes of dark matter haloes on a scale
difficult to probe with weak lensing or any other mass estimator.  The
caustic pattern is often visible up to and beyond the turnaround
radius.  The mass within the turnaround radius (for clusters where the
radial extent of the caustics $r_{max}$ exceeds the turnaround radius)
is 2.19$\pm$0.18 times the virial mass $M_{200}$ (the average ratio
for all clusters is 1.97$\pm$0.10), in agreement with the numerical
simulations of $\Lambda$CDM clusters of \citet{busha05}, who find that
the final mass of cluster scale halos in the far future is 1.9 times
larger than $M_{200}$ measured at the present epoch.

We stack the clusters to produce an ensemble cluster containing
4502 galaxies projected within $r_{200}$ and an additional 10,601 within
5$r_{200}$ (roughly the turnaround radius). The infall region thus
contains more galaxies than the virial region. The ensemble cluster
appears circularly symmetric. 

At small radii, the caustic mass profiles are consistent with
independent X-ray mass estimates using previously determined scaling
relations such as those found for RASS-SDSS \citep{popesso05}.  This
good agreement indicates that the caustic technique is a reliable mass
estimator \citep[see also][]{cairnsi}.  At larger radii, the caustic
masses agree well with virial masses.

The CAIRNS project demonstrated that the caustic pattern is common in
rich, X-ray luminous galaxy clusters.  The CIRS sample, eight times
larger than the CAIRNS sample, confirms and extends many of the
results of CAIRNS.  Future papers in the CIRS project will analyze the
relative distributions of mass and light in cluster infall regions,
X-ray and optical substructure within infall regions, and the
dependence of galaxy properties on environment.

\acknowledgements

We thank Margaret Geller for useful discussions during early stages of
this work and for insightful comments on early drafts.  We thank the
Smithsonian Astrophysical Observatory for use of their computer
facilities for some computations.  KR was a visitor at SAO for part of
this project and thanks them for their hospitality.  Funding for the
Sloan Digital Sky Survey (SDSS) has been provided by the Alfred
P. Sloan Foundation, the Participating Institutions, the National
Aeronautics and Space Administration, the National Science Foundation,
the U.S. Department of Energy, the Japanese Monbukagakusho, and the
Max Planck Society. The SDSS Web site is http://www.sdss.org/.  The
SDSS is managed by the Astrophysical Research Consortium (ARC) for the
Participating Institutions. The Participating Institutions are The
University of Chicago, Fermilab, the Institute for Advanced Study, the
Japan Participation Group, The Johns Hopkins University, the Korean
Scientist Group, Los Alamos National Laboratory, the
Max-Planck-Institute for Astronomy (MPIA), the Max-Planck-Institute
for Astrophysics (MPA), New Mexico State University, University of
Pittsburgh, University of Portsmouth, Princeton University, the United
States Naval Observatory, and the University of Washington.  We thank
the referee for helpful comments that improved the presentation of
this paper.

\bibliographystyle{apj}
\bibliography{rines}

\begin{thebibliography}{95}
\expandafter\ifx\csname natexlab\endcsname\relax\def\natexlab#1{#1}\fi

\bibitem[{{Adelman-McCarthy} {et~al.}(2006)}]{dr4}
{Adelman-McCarthy}, J. {et~al.} 2006, \apjs, 162, 38

\bibitem[{{B{\" o}hringer} {et~al.}(2000{\natexlab{a}})}]{2000ApJS..129..435B}
{B{\" o}hringer}, H. {et~al.} 2000{\natexlab{a}}, \apjs, 129, 435

\bibitem[{{B{\" o}hringer} {et~al.}(2000{\natexlab{b}})}]{noras}
---. 2000{\natexlab{b}}, \apjs, 129, 435

\bibitem[{{B{\" o}hringer} {et~al.}(2001)}]{2001A&A...369..826B}
---. 2001, \aap, 369, 826

\bibitem[{{B{\" o}hringer} {et~al.}(2004)}]{2004A&A...425..367B}
---. 2004, \aap, 425, 367

\bibitem[{{Belsole} {et~al.}(2004){Belsole}, {Pratt}, {Sauvageot}, \&
  {Bourdin}}]{belsole04}
{Belsole}, E., {Pratt}, G.~W., {Sauvageot}, J.-L., \& {Bourdin}, H. 2004, \aap,
  415, 821

\bibitem[{{Berlind} {et~al.}(2003)}]{berlind03}
{Berlind}, A.~A. {et~al.} 2003, \apj, 593, 1

\bibitem[{{Biviano} \& {Girardi}(2003)}]{bg03}
{Biviano}, A. \& {Girardi}, M. 2003, \apj, 585, 205

\bibitem[{{Blanton} {et~al.}(2003)}]{blanton03}
{Blanton}, M.~R. {et~al.} 2003, \apj, 592, 819

\bibitem[{{Borgani} {et~al.}(2004)}]{borgani04}
{Borgani}, S. {et~al.} 2004, \mnras, 348, 1078

\bibitem[{{Borgani} {et~al.}(2006)}]{borgani06}
---. 2006, \mnras, 270

\bibitem[{{Broadhurst} {et~al.}(2005{\natexlab{a}}){Broadhurst}, {Takada},
  {Umetsu}, {Kong}, {Arimoto}, {Chiba}, \& {Futamase}}]{2005ApJ...619L.143B}
{Broadhurst}, T., {Takada}, M., {Umetsu}, K., {Kong}, X., {Arimoto}, N.,
  {Chiba}, M., \& {Futamase}, T. 2005{\natexlab{a}}, \apjl, 619, L143

\bibitem[{{Broadhurst} {et~al.}(2005{\natexlab{b}})}]{2005ApJ...621...53B}
{Broadhurst}, T. {et~al.} 2005{\natexlab{b}}, \apj, 621, 53

\bibitem[{{Bullock} {et~al.}(2001)}]{bullock01}
{Bullock}, J.~S. {et~al.} 2001, \mnras, 321, 559

\bibitem[{{Busha} {et~al.}(2003){Busha}, {Adams}, {Wechsler}, \&
  {Evrard}}]{busha03}
{Busha}, M.~T., {Adams}, F.~C., {Wechsler}, R.~H., \& {Evrard}, A.~E. 2003,
  \apj, 596, 713

\bibitem[{{Busha} {et~al.}(2005){Busha}, {Evrard}, {Adams}, \&
  {Wechsler}}]{busha05}
{Busha}, M.~T., {Evrard}, A.~E., {Adams}, F.~C., \& {Wechsler}, R.~H. 2005,
  \mnras, 363, L11

\bibitem[{{Carlberg} {et~al.}(1997){Carlberg}, {Yee}, \& {Ellingson}}]{cye97}
{Carlberg}, R.~G., {Yee}, H.~K.~C., \& {Ellingson}, E. 1997, \apj, 478, 462

\bibitem[{{Cavaliere} \& {Fusco-Femiano}(1976)}]{cff}
{Cavaliere}, A. \& {Fusco-Femiano}, R. 1976, \aap, 49, 137

\bibitem[{{Clowe} \& {Schneider}(2001)}]{clowe01}
{Clowe}, D. \& {Schneider}, P. 2001, \aap, 379, 384

\bibitem[{{Col{\' i}n} {et~al.}(1999){Col{\' i}n}, {Klypin}, {Kravtsov}, \&
  {Khokhlov}}]{1999ApJ...523...32C}
{Col{\' i}n}, P., {Klypin}, A.~A., {Kravtsov}, A.~V., \& {Khokhlov}, A.~M.
  1999, \apj, 523, 32

\bibitem[{{Cross} {et~al.}(2004){Cross}, {Driver}, {Liske}, {Lemon}, {Peacock},
  {Cole}, {Norberg}, \& {Sutherland}}]{2004MNRAS.349..576C}
{Cross}, N.~J.~G., {Driver}, S.~P., {Liske}, J., {Lemon}, D.~J., {Peacock},
  J.~A., {Cole}, S., {Norberg}, P., \& {Sutherland}, W.~J. 2004, \mnras, 349,
  576

\bibitem[{{Danese} {et~al.}(1980){Danese}, {de Zotti}, \& {di Tullio}}]{danese}
{Danese}, L., {de Zotti}, G., \& {di Tullio}, G. 1980, \aap, 82, 322

\bibitem[{{David} {et~al.}(1993){David}, {Slyz}, {Jones}, {Forman}, {Vrtilek},
  \& {Arnaud}}]{david93}
{David}, L.~P., {Slyz}, A., {Jones}, C., {Forman}, W., {Vrtilek}, S.~D., \&
  {Arnaud}, K.~A. 1993, \apj, 412, 479

\bibitem[{{Diaferio}(1999)}]{diaferio1999}
{Diaferio}, A. 1999, \mnras, 309, 610

\bibitem[{{Diaferio} \& {Geller}(1997)}]{dg97}
{Diaferio}, A. \& {Geller}, M.~J. 1997, \apj, 481, 633

\bibitem[{{Diaferio} {et~al.}(2005){Diaferio}, {Geller}, \&
  {Rines}}]{diaferio05}
{Diaferio}, A., {Geller}, M.~J., \& {Rines}, K.~J. 2005, \apjl, 628, L97

\bibitem[{{Diemand} {et~al.}(2004){Diemand}, {Moore}, \& {Stadel}}]{diemand04}
{Diemand}, J., {Moore}, B., \& {Stadel}, J. 2004, \mnras, 352, 535

\bibitem[{{Drinkwater} {et~al.}(2001){Drinkwater}, {Gregg}, \&
  {Colless}}]{drink}
{Drinkwater}, M.~J., {Gregg}, M.~D., \& {Colless}, M. 2001, \apj, 548, L139

\bibitem[{{Ebeling} {et~al.}(2000{\natexlab{a}}){Ebeling}, {Edge}, {Allen},
  {Crawford}, {Fabian}, \& {Huchra}}]{2000MNRAS.318..333E}
{Ebeling}, H., {Edge}, A.~C., {Allen}, S.~W., {Crawford}, C.~S., {Fabian},
  A.~C., \& {Huchra}, J.~P. 2000{\natexlab{a}}, \mnras, 318, 333

\bibitem[{{Ebeling} {et~al.}(2000{\natexlab{b}}){Ebeling}, {Edge}, {Allen},
  {Crawford}, {Fabian}, \& {Huchra}}]{bcs}
---. 2000{\natexlab{b}}, \mnras, 318, 333

\bibitem[{{Ebeling} {et~al.}(1996){Ebeling}, {Voges}, {Bohringer}, {Edge},
  {Huchra}, \& {Briel}}]{xbacs}
{Ebeling}, H., {Voges}, W., {Bohringer}, H., {Edge}, A.~C., {Huchra}, J.~P., \&
  {Briel}, U.~G. 1996, \mnras, 281, 799

\bibitem[{{Eke} {et~al.}(1996){Eke}, {Cole}, \& {Frenk}}]{ecf96}
{Eke}, V.~R., {Cole}, S., \& {Frenk}, C.~S. 1996, \mnras, 282, 263

\bibitem[{{Ellingson} {et~al.}(2001){Ellingson}, {Lin}, {Yee}, \&
  {Carlberg}}]{ellingson01}
{Ellingson}, E., {Lin}, H., {Yee}, H. K.~C., \& {Carlberg}, R.~G. 2001, \apj,
  547, 609

\bibitem[{{Evrard} {et~al.}(1996){Evrard}, {Metzler}, \& {Navarro}}]{emn96}
{Evrard}, A.~E., {Metzler}, C.~A., \& {Navarro}, J.~F. 1996, \apj, 469, 494

\bibitem[{{Fadda} {et~al.}(1996){Fadda}, {Girardi}, {Giuricin}, {Mardirossian},
  \& {Mezzetti}}]{1996ApJ...473..670F}
{Fadda}, D., {Girardi}, M., {Giuricin}, G., {Mardirossian}, F., \& {Mezzetti},
  M. 1996, \apj, 473, 670

\bibitem[{{Faltenbacher} {et~al.}(2005){Faltenbacher}, {Kravtsov}, {Nagai}, \&
  {Gottl{\"o}ber}}]{faltenbacher05}
{Faltenbacher}, A., {Kravtsov}, A.~V., {Nagai}, D., \& {Gottl{\"o}ber}, S.
  2005, \mnras, 358, 139

\bibitem[{{Faltenbacher} {et~al.}(2006){Faltenbacher}, {Kravtsov}, {Nagai}, \&
  {Gottl{\"o}ber}}]{faltenbacher06}
---. 2006, \mnras, submitted (astro-ph/0602197)

\bibitem[{{Finoguenov} {et~al.}(2001){Finoguenov}, {Reiprich}, \& {B{\"
  o}hringer}}]{frb2001}
{Finoguenov}, A., {Reiprich}, T.~H., \& {B{\" o}hringer}, H. 2001, \aap, 368,
  749

\bibitem[{{Gavazzi}(2005)}]{gavazzi05}
{Gavazzi}, R. 2005, \aap, 443, 793

\bibitem[{{Geller} {et~al.}(1999){Geller}, {Diaferio}, \& {Kurtz}}]{gdk99}
{Geller}, M.~J., {Diaferio}, A., \& {Kurtz}, M.~J. 1999, \apjl, 517, L23

\bibitem[{{Girardi} {et~al.}(1998){Girardi}, {Giuricin}, {Mardirossian},
  {Mezzetti}, \& {Boschin}}]{girardi98}
{Girardi}, M., {Giuricin}, G., {Mardirossian}, F., {Mezzetti}, M., \&
  {Boschin}, W. 1998, \apj, 505, 74

\bibitem[{{Gramann} \& {Suhhonenko}(2002)}]{2002MNRAS.337.1417G}
{Gramann}, M. \& {Suhhonenko}, I. 2002, \mnras, 337, 1417

\bibitem[{{Heisler} {et~al.}(1985){Heisler}, {Tremaine}, \& {Bahcall}}]{htb}
{Heisler}, J., {Tremaine}, S., \& {Bahcall}, J.~N. 1985, \apj, 298, 8

\bibitem[{{Hernquist}(1990)}]{hernquist1990}
{Hernquist}, L. 1990, \apj, 356, 359

\bibitem[{{Horner}(2001)}]{hornerthesis}
{Horner}, D. 2001, Ph.D.~Thesis, University of Maryland

\bibitem[{{Horner} {et~al.}(1999){Horner}, {Mushotzky}, \&
  {Scharf}}]{1999ApJ...520...78H}
{Horner}, D.~J., {Mushotzky}, R.~F., \& {Scharf}, C.~A. 1999, \apj, 520, 78

\bibitem[{{Jones} \& {Forman}(1999)}]{jf99}
{Jones}, C. \& {Forman}, W. 1999, \apj, 511, 65

\bibitem[{{Kaiser}(1987)}]{kais87}
{Kaiser}, N. 1987, \mnras, 227, 1

\bibitem[{{Kauffmann} {et~al.}(1999{\natexlab{a}}){Kauffmann}, {Colberg},
  {Diaferio}, \& {White}}]{kauffmann1999a}
{Kauffmann}, G., {Colberg}, J.~M., {Diaferio}, A., \& {White}, S. D.~M.
  1999{\natexlab{a}}, \mnras, 303, 188

\bibitem[{{Kauffmann} {et~al.}(1999{\natexlab{b}}){Kauffmann}, {Colberg},
  {Diaferio}, \& {White}}]{kauffmann1999b}
---. 1999{\natexlab{b}}, \mnras, 307, 529

\bibitem[{{Klypin} {et~al.}(1999){Klypin}, {Gottl{\" o}ber}, {Kravtsov}, \&
  {Khokhlov}}]{1999ApJ...516..530K}
{Klypin}, A., {Gottl{\" o}ber}, S., {Kravtsov}, A.~V., \& {Khokhlov}, A.~M.
  1999, \apj, 516, 530

\bibitem[{{Kneib} {et~al.}(2003)}]{kneib03}
{Kneib}, J.-P. {et~al.} 2003, \apj, 598, 804

\bibitem[{{Koranyi}(2000)}]{dkthesis}
{Koranyi}, D.~M. 2000, Ph.D.~Thesis, Harvard University

\bibitem[{{Koranyi} \& {Geller}(2000)}]{kg2000}
{Koranyi}, D.~M. \& {Geller}, M.~J. 2000, \aj, 119, 44

\bibitem[{{Lahav} {et~al.}(1991){Lahav}, {Rees}, {Lilje}, \&
  {Primack}}]{1991MNRAS.251..128L}
{Lahav}, O., {Rees}, M.~J., {Lilje}, P.~B., \& {Primack}, J.~R. 1991, \mnras,
  251, 128

\bibitem[{{Ledlow} {et~al.}(2003){Ledlow}, {Voges}, {Owen}, \&
  {Burns}}]{2003AJ....126.2740L}
{Ledlow}, M.~J., {Voges}, W., {Owen}, F.~N., \& {Burns}, J.~O. 2003, \aj, 126,
  2740

\bibitem[{{Lin} {et~al.}(2004){Lin}, {Mohr}, \& {Stanford}}]{lin04}
{Lin}, Y., {Mohr}, J.~J., \& {Stanford}, S.~A. 2004, \apj, 610, 745

\bibitem[{{{\L}okas} \& {Mamon}(2003)}]{lokas03}
{{\L}okas}, E.~L. \& {Mamon}, G.~A. 2003, \mnras, 343, 401

\bibitem[{{Mahdavi} {et~al.}(1999){Mahdavi}, {Geller}, {B{\" o}hringer},
  {Kurtz}, \& {Ramella}}]{mahdavi99}
{Mahdavi}, A., {Geller}, M.~J., {B{\" o}hringer}, H., {Kurtz}, M.~J., \&
  {Ramella}, M. 1999, \apj, 518, 69

\bibitem[{{Mahdavi} {et~al.}(2005){Mahdavi}, {Trentham}, \&
  {Tully}}]{mahdavi05}
{Mahdavi}, A., {Trentham}, N., \& {Tully}, R.~B. 2005, \aj, 130, 1502

\bibitem[{{Majumdar} \& {Mohr}(2004)}]{majumdar04}
{Majumdar}, S. \& {Mohr}, J.~J. 2004, \apj, 613, 41

\bibitem[{{Marini} {et~al.}(2004)}]{marini04}
{Marini}, F. {et~al.} 2004, \mnras, 353, 1219

\bibitem[{{Miller} {et~al.}(2005)}]{miller05}
{Miller}, C.~J. {et~al.} 2005, \aj, 130, 968

\bibitem[{{Moore} {et~al.}(1999){Moore}, {Quinn}, {Governato}, {Stadel}, \&
  {Lake}}]{moore99}
{Moore}, B., {Quinn}, T., {Governato}, F., {Stadel}, J., \& {Lake}, G. 1999,
  \mnras, 310, 1147

\bibitem[{{Nagamine} \& {Loeb}(2003)}]{nl02}
{Nagamine}, K. \& {Loeb}, A. 2003, New Astronomy, 8, 439

\bibitem[{{Navarro} {et~al.}(1997){Navarro}, {Frenk}, \& {White}}]{nfw97}
{Navarro}, J.~F., {Frenk}, C.~S., \& {White}, S. D.~M. 1997, \apj, 490, 493

\bibitem[{{Nevalainen} {et~al.}(2000){Nevalainen}, {Markevitch}, \&
  {Forman}}]{2000ApJ...532..694N}
{Nevalainen}, J., {Markevitch}, M., \& {Forman}, W. 2000, \apj, 532, 694

\bibitem[{{Plaga}(2005)}]{2005A&A...440L..41P}
{Plaga}, R. 2005, \aap, 440, L41

\bibitem[{{Popesso} {et~al.}(2005){Popesso}, {Biviano}, {B{\"o}hringer},
  {Romaniello}, \& {Voges}}]{popesso05}
{Popesso}, P., {Biviano}, A., {B{\"o}hringer}, H., {Romaniello}, M., \&
  {Voges}, W. 2005, \aap, 433, 431

\bibitem[{{Popesso} {et~al.}(2004){Popesso}, {B{\"o}hringer}, {Brinkmann},
  {Voges}, \& {York}}]{popesso04}
{Popesso}, P., {B{\"o}hringer}, H., {Brinkmann}, J., {Voges}, W., \& {York},
  D.~G. 2004, \aap, 423, 449

\bibitem[{{Postman} {et~al.}(1988){Postman}, {Geller}, \&
  {Huchra}}]{1988AJ.....95..267P}
{Postman}, M., {Geller}, M.~J., \& {Huchra}, J.~P. 1988, \aj, 95, 267

\bibitem[{{Reg\"os} \& {Geller}(1989)}]{rg89}
{Reg\"os}, E. \& {Geller}, M.~J. 1989, \aj, 98, 755

\bibitem[{{Reisenegger} {et~al.}(2000){Reisenegger}, {Quintana}, {Carrasco}, \&
  {Maze}}]{rqcm}
{Reisenegger}, A., {Quintana}, H., {Carrasco}, E.~R., \& {Maze}, J. 2000, \aj,
  120, 523

\bibitem[{{Rines} {et~al.}(2004){Rines}, {Geller}, {Diaferio}, {Kurtz}, \&
  {Jarrett}}]{cairnsii}
{Rines}, K., {Geller}, M.~J., {Diaferio}, A., {Kurtz}, M.~J., \& {Jarrett},
  T.~H. 2004, \aj, 128, 1078

\bibitem[{{Rines} {et~al.}(2002){Rines}, {Geller}, {Diaferio}, {Mahdavi},
  {Mohr}, \& {Wegner}}]{rines02}
{Rines}, K., {Geller}, M.~J., {Diaferio}, A., {Mahdavi}, A., {Mohr}, J.~J., \&
  {Wegner}, G. 2002, \aj, 124, 1266

\bibitem[{{Rines} {et~al.}(2000){Rines}, {Geller}, {Diaferio}, {Mohr}, \&
  {Wegner}}]{rines2000}
{Rines}, K., {Geller}, M.~J., {Diaferio}, A., {Mohr}, J.~J., \& {Wegner}, G.~A.
  2000, \aj, 120, 2338

\bibitem[{{Rines} {et~al.}(2003){Rines}, {Geller}, {Kurtz}, \&
  {Diaferio}}]{cairnsi}
{Rines}, K., {Geller}, M.~J., {Kurtz}, M.~J., \& {Diaferio}, A. 2003, \aj, 126,
  2152

\bibitem[{{Rines} {et~al.}(2005){Rines}, {Geller}, {Kurtz}, \&
  {Diaferio}}]{cairnsha}
---. 2005, \aj, 130, 1482

\bibitem[{{Rines} {et~al.}(2001){Rines}, {Mahdavi}, {Geller}, {Diaferio},
  {Mohr}, \& {Wegner}}]{rines01b}
{Rines}, K., {Mahdavi}, A., {Geller}, M.~J., {Diaferio}, A., {Mohr}, J.~J., \&
  {Wegner}, G. 2001, \apj, 555, 558

\bibitem[{{Slosar} {et~al.}(2006){Slosar}, {Seljak}, \&
  {Tasitsiomi}}]{2006MNRAS.366.1455S}
{Slosar}, A., {Seljak}, U., \& {Tasitsiomi}, A. 2006, \mnras, 366, 1455

\bibitem[{{Small} {et~al.}(1998){Small}, {Ma}, {Sargent}, \&
  {Hamilton}}]{small98}
{Small}, T.~A., {Ma}, C., {Sargent}, W. L.~W., \& {Hamilton}, D. 1998, \apj,
  492, 45

\bibitem[{{Springel} {et~al.}(2005)}]{springel05}
{Springel}, V. {et~al.} 2005, \nat, 435, 629

\bibitem[{{Stoughton} {et~al.}(2002)}]{sdss}
{Stoughton}, C. {et~al.} 2002, \aj, 123, 485

\bibitem[{{Strauss} {et~al.}(2002)}]{strauss02}
{Strauss}, M.~A. {et~al.} 2002, \aj, 124, 1810

\bibitem[{{Tegmark} {et~al.}(2004)}]{tegmark04}
{Tegmark}, M. {et~al.} 2004, \prd, 69, 103501

\bibitem[{{The} \& {White}(1986)}]{1986AJ.....92.1248T}
{The}, L.~S. \& {White}, S.~D.~M. 1986, \aj, 92, 1248

\bibitem[{{Tinker} {et~al.}(2005){Tinker}, {Weinberg}, {Zheng}, \&
  {Zehavi}}]{tinker05}
{Tinker}, J.~L., {Weinberg}, D.~H., {Zheng}, Z., \& {Zehavi}, I. 2005, \apj,
  631, 41

\bibitem[{{van den Bosch} {et~al.}(2003){van den Bosch}, {Mo}, \&
  {Yang}}]{2003MNRAS.345..923V}
{van den Bosch}, F.~C., {Mo}, H.~J., \& {Yang}, X. 2003, \mnras, 345, 923

\bibitem[{{van den Bosch} {et~al.}(2005){van den Bosch}, {Weinmann}, {Yang},
  {Mo}, {Li}, \& {Jing}}]{2005MNRAS.361.1203V}
{van den Bosch}, F.~C., {Weinmann}, S.~M., {Yang}, X., {Mo}, H.~J., {Li}, C.,
  \& {Jing}, Y.~P. 2005, \mnras, 361, 1203

\bibitem[{{van der Marel} {et~al.}(2000){van der Marel}, {Magorrian},
  {Carlberg}, {Yee}, \& {Ellingson}}]{2000AJ....119.2038V}
{van der Marel}, R.~P., {Magorrian}, J., {Carlberg}, R.~G., {Yee}, H.~K.~C., \&
  {Ellingson}, E. 2000, \aj, 119, 2038

\bibitem[{{Vedel} \& {Hartwick}(1998)}]{vh98}
{Vedel}, H. \& {Hartwick}, F.~D.~A. 1998, \apj, 501, 509

\bibitem[{{Voges} {et~al.}(1999)}]{rass}
{Voges}, W. {et~al.} 1999, \aap, 349, 389

\bibitem[{{Yoshikawa} {et~al.}(2003){Yoshikawa}, {Jing}, \&
  {B{\"o}rner}}]{2003ApJ...590..654Y}
{Yoshikawa}, K., {Jing}, Y.~P., \& {B{\"o}rner}, G. 2003, \apj, 590, 654

\bibitem[{{Zwicky}(1933)}]{zwicky1933}
{Zwicky}, F. 1933, Helv.~Phys.~Acta, 6, 110

\bibitem[{{Zwicky}(1937)}]{zwicky1937}
---. 1937, \apj, 86, 217

\end{thebibliography}

\clearpage

\clearpage

%\begin{figure}
%\figurenum{25}
%\plotone{f25.eps}
%\caption{\label{allcirss2} See Figure \ref{allcirss1}.}  
%\end{figure}

%\begin{figure}
%\figurenum{26}
%\plotone{f26.eps}
%\caption{\label{allcirss3} See Figure \ref{allcirss1}.}  
%\end{figure}

%\begin{figure}
%\figurenum{27}
%\plotone{f27.eps}
%\caption{\label{allcirss4} See Figure \ref{allcirss1}.}  
%\end{figure}

%\begin{figure}
%\figurenum{28}
%\plotone{f28.eps}
%\caption{\label{allcirss5} See Figure \ref{allcirss1}.}  
%\end{figure}

%\begin{figure}
%\figurenum{29}
%\plotone{f29.eps}
%\caption{\label{allcirss6} See Figure \ref{allcirss1}.}  
%\end{figure}

%\begin{figure}
%\figurenum{30}
%%\epsscale{1.0}
%%\plotone{Rines.fig7.eps} 
%\caption{\label{combovdp} Velocity dispersion profile of the ensemble CAIRNS
%cluster. The filled squares and solid line show the velocity
%dispersion profile of member galaxies (those within the caustics) with
%1-$\sigma$ uncertainties.  The open triangles and dashed line show the
%enclosed velocity dispersion.  The dash-dotted line shows the VDP
%predicted by the Hernquist mass models which best fits the caustic
%mass profiles (assuming isotropic orbits).  The inset shows a closeup
%of the VDP within $r_{200}$.}
%\end{figure}

\clearpage

\begin{table*}[th] \footnotesize
\begin{center}
\caption{\label{sample} \sc CIRS Basic Properties}
\begin{tabular}{lcccrrcccc}
\tableline
\tableline
\tablewidth{0pt}
Cluster &\multicolumn{2}{c}{X-ray Coordinates} & $z_\odot$ & $L_X /10^{43}$ & Catalog & $T_X$ & $\sigma_p$ & Flag & $R_{comp}$ \\ 
 & RA (J2000) & DEC (J2000) &   & erg~s$^{-1}$& & keV  & $\kms$ & & $\Mpc$ \\ 
\tableline
      A0085 & 10.45880 & -9.30190 & 0.0557 & 2.805 &  REF & 5.87 & $ 692 ^{+55} _{-45}$  & 2 & 1.8  \\ 
      A0119 & 14.07620 & -1.21670 & 0.0446 & 0.781 &  REF & 5.93 & $ 589 ^{+68} _{-51}$  & 2 & 0.1  \\ 
      A0160 & 18.27410 & 15.51700 & 0.0432 & 0.092 &  NOR & 2.70 & $ 489 ^{+62} _{-45}$  & 2 & 0.7  \\ 
      A0168 & 18.80000 & 00.33000 & 0.0451 & 0.247 &  REF & 2.60 & $ 577 ^{+50} _{-40}$  & 2 & 2.1  \\ 
    RXJ0137 & 24.31420 & -9.20280 & 0.0409 & 0.155 &  REF & 0.00 & $ 392 ^{+45} _{-34}$  & 2 & 2.1  \\ 
\tableline
\tablenotetext{a}{X-ray luminosity assuming all X-ray flux due to this component.}
%\tablenotetext{b}{Zabludoff et al.~1993}
\end{tabular}
\tablecomments{
Table \ref{sample} is published in its entirety in the electronic
edition of the Astronomical Journal.  A portion is shown here for
guidance regarding its form and content.}
\end{center}
\end{table*}

\begin{table*}[th] \footnotesize
\begin{center}
\caption{\label{centers} \sc CIRS Hierarchical Centers and Offsets}
\begin{tabular}{lcccr}
\tableline
\tableline
\tablewidth{0pt}
Cluster &\multicolumn{2}{c}{Hierarchical Center} & $\Delta cz$ & $\Delta R$  \\ 
 & RA (J2000) & DEC (J2000) & $\kms$ & $\kpc$  \\ 
\tableline
      A0085 & 10.42629 & -9.42550 & -45 & 347   \\ 
      A0119 & 14.03449 & -1.16356 & -106 & 149   \\ 
      A0160 & 18.25117 & 15.49587 & 285 & 65   \\ 
      A0168 & 18.81422 & 00.26408 & -17 & 150   \\ 
    RXJ0137 & 24.35443 & -9.27309 & -12 & 164   \\ 
\tableline
\end{tabular}
\tablecomments{
Table \ref{centers} is published in its entirety in the electronic
edition of the Astronomical Journal.  A portion is shown here for
guidance regarding its form and content.}
\end{center}
\end{table*}

\begin{table*}[th] \footnotesize
\begin{center}
\caption{\label{radii} \sc CIRS Characteristic Radii and Masses}
\begin{tabular}{lccrrrrr}
\tableline
\tableline
\tablewidth{0pt}
Cluster & $r_{500}$ & $r_{200}$ & $r_t$ & $r_{max}$ & $M_{200}$ & $M_{t}$ & $M_{max}/M_{200}$ \\ 
 & $\Mpc$ & $\Mpc$ & $\Mpc$ & $\Mpc$ & $10^{14} M_\odot$ & $10^{14} M_\odot$ &  \\ 
\tableline
      A0085 & 0.67 & 1.02 & 4.99 & 4.34 & 2.50$\pm$ 1.19 & 5.04$\pm$2.95 & 2.02   \\ 
      A0119 & 0.60 & 0.91 & 4.69 & 3.64 & 1.77$\pm$ 0.70 & 4.19$\pm$1.88 & 2.37   \\ 
      A0160 & 0.46 & 0.67 & 2.70 & 1.31 & 0.68$\pm$ 0.38 & 0.80$\pm$0.54 & 1.17   \\ 
      A0168 & 0.63 & 0.95 & 4.19 & 3.94 & 2.02$\pm$ 0.51 & 3.00$\pm$0.93 & 1.48   \\ 
    RXJ0137 & 0.46 & 0.72 & 3.14 & 2.42 & 0.87$\pm$ 0.06 & 1.26$\pm$0.09 & 1.45   \\ 
\tableline
\end{tabular}
\tablecomments{
Table \ref{radii} is published in its entirety in the electronic
edition of the Astronomical Journal.  A portion is shown here for
guidance regarding its form and content.}
\end{center}
\end{table*}

\begin{table*}[th] \footnotesize
\begin{center}
\caption{\label{centerproblems} \sc Individual Limits for CIRS Clusters }
\begin{tabular}{lcc}
\tableline
\tableline
\tablewidth{0pt}
Cluster & $R_p$ & $\Delta cz$  \\ 
 & $\Mpc$  & $\kms$ \\ 
\tableline
A160  & -- & $-1500\rightarrow +3000$ \\
RXJ0137  & 5 &  $-5000\rightarrow +3000$ \\
A954  & 5 & --  \\
A1035B & -- & $+1500\rightarrow +7000$\\
A1173  & 3 & --  \\
A1291A & 5 &$-5000\rightarrow -1000$ \\
A1377 & 5 & $-5000\rightarrow +2000$ \\
A1436 & 6 & -- \\
RXJ1210  & 4 & --  \\ 
NGC4325 & 1.5 & -- \\
NGC4636 & 1.0 & $\pm$2000 \\
RXJ1351  & 3 & --  \\
MS1306  & 5 & --  \\
NGC5846 & 5 & -- \\
A2067 & 1.5 & --  \\ 
A2149  & 4 & --  \\
NGC6107 & 4 & --  \\
A2197 & 1.5 & -- \\
A2245 & -- & $-5000\rightarrow +3000$ \\
A2244 & 1.5 & $-3000\rightarrow +5000$ \\
A2249 & 6 & $\pm 4000$ \\
\tableline
\end{tabular}
\end{center}
\end{table*}

\begin{table*}[th] \footnotesize
\begin{center}
\caption{\label{mpfitsci} \sc CIRS Mass Profile Fit Parameters}
\begin{tabular}{lcccccc}
\tableline
\tableline
\tablewidth{0pt}
Cluster & $a_{NFW}$ & $r_{200}$ & $c_{NFW}$ & $M_{200}$ & Best-fit & $c_{101}$\\ 
 & $\Mpc$ & $\Mpc$ & & $10^{14} M_\odot$ &  Profile \\ 
\tableline
        A85 & 0.223 & 1.00 & 4.50 & 2.35 & H & 6.07   \\ 
       A119 & 0.350 & 0.89 & 2.55 & 1.65 & H & 3.52   \\ 
      A160 & 0.064 & 0.65 & 10.14 & 0.64 & H & 13.36   \\ 
       A168 & 0.114 & 0.88 & 7.69 & 1.57 & H & 10.20   \\ 
   RXJ0137 & 0.105 & 0.67 & 6.34 & 0.69 & H & 8.45   \\ 
\tableline
\end{tabular}
\tablecomments{
Table \ref{mpfitsci} is published in its entirety in the electronic
edition of the Astronomical Journal.  A portion is shown here for
guidance regarding its form and content.}
\end{center}
\end{table*}

\begin{table*}[th] \footnotesize
\begin{center}
\caption{\label{virial} \sc CIRS Virial and Projected Masses}
\begin{tabular}{lcrrr}
\tableline
\tableline
\tablewidth{0pt}
Cluster & $r_{200}$ & $M_{200}$ & $M_{proj}$ & $M_{vir}$  \\ 
 & $\Mpc$  & $10^{14} M_\odot$ & $10^{14} M_\odot$   & $10^{14} M_\odot$  \\ 
\tableline
      A0085 & 1.02 & 2.50$\pm$ 1.19 & 2.11$\pm$0.30 & 2.94$\pm$0.29   \\ 
      A0119 & 0.91 & 1.77$\pm$ 0.70 & 1.68$\pm$0.33 & 1.33$\pm$0.17   \\ 
      A0160 & 0.67 & 0.68$\pm$ 0.38 & 0.56$\pm$0.12 & 1.17$\pm$0.15   \\ 
      A0168 & 0.95 & 2.02$\pm$ 0.51 & 1.49$\pm$0.23 & 2.02$\pm$0.21   \\ 
    RXJ0137 & 0.72 & 0.87$\pm$ 0.06 & 0.63$\pm$0.12 & 0.84$\pm$0.10   \\ 
\tableline
\end{tabular}
\tablecomments{
Table \ref{virial} is published in its entirety in the electronic
edition of the Astronomical Journal.  A portion is shown here for
guidance regarding its form and content.}
\end{center}
\end{table*}

\end{document}